\documentclass[10pt,aps,showpacs,prx,superscriptaddress,preprintnumbers,amssymb,twocolumn]{revtex4-2}

\usepackage{amsmath}
\usepackage{placeins}
\usepackage[utf8]{inputenc}
\usepackage[T1]{fontenc}
\usepackage{graphicx}
\usepackage{dcolumn}
\usepackage{bm}
\usepackage[dvipsnames]{xcolor}
\usepackage{braket}
\usepackage{bbold}
\usepackage{mathtools}
\usepackage{dsfont}
\usepackage{siunitx}
\usepackage{caption}

\usepackage{subfig}
\usepackage{bbm}
\usepackage{enumerate}

\usepackage{hyperref}
\hypersetup{
    colorlinks=true,
    linkcolor=blue,
    filecolor=magenta,      
    urlcolor=cyan
}
\usepackage{titlesec}
\usepackage{array}
\usepackage{physics}

\usepackage{changes}

\definecolor{blue-violet}{rgb}{0.54, 0.17, 0.89}

\makeatletter
\def\@fnsymbol#1{\ensuremath{\ifcase#1\or \dagger\or *\or
   \mathsection\or \mathparagraph\or \|\or **\or \dagger\dagger
   \or ** \else\@ctrerr\fi}}

\begin{document}

\allowdisplaybreaks

\title{Long-distance device-independent quantum key distribution using single-photon entanglement}

\author{Anna Steffinlongo}
\thanks{These authors contributed equally to this work}
\affiliation{ICFO - Institut de Ciencies Fotoniques, The Barcelona Institute of Science and Technology, 08860 Castelldefels, Barcelona, Spain}

\author{Mariana Navarro}
\thanks{These authors contributed equally to this work}
\affiliation{ICFO - Institut de Ciencies Fotoniques, The Barcelona Institute of Science and Technology, 08860 Castelldefels, Barcelona, Spain}
\affiliation{LuxQuanta Technologies S.L., Mediterranean Technology Park. Carrer d’Esteve Terradas, 1, Office 206, 08860 Castelldefels, Barcelona, Spain}

\author{Marina Cenni}
\affiliation{ICFO - Institut de Ciencies Fotoniques, The Barcelona Institute of Science and Technology, 08860 Castelldefels, Barcelona, Spain}

\author{Xavier Valcarce}
\affiliation{Université Paris-Saclay, CEA, CNRS, Institut de physique théorique, 91191, Gif-sur-Yvette, France}
\author{Antonio Acín}
\affiliation{ICFO - Institut de Ciencies Fotoniques, The Barcelona Institute of Science and Technology, 08860 Castelldefels, Barcelona, Spain}
\affiliation{ICREA – Institució Catalana de Recerca i Estudis Avançats, Lluís Companys 23, 08010 Barcelona, Spain}

\author{Enky Oudot}
\email{enky.oudot@lip6.fr}
\affiliation{ICFO - Institut de Ciencies Fotoniques, The Barcelona Institute of Science and Technology, 08860 Castelldefels, Barcelona, Spain}
\affiliation{LIP6, CNRS, Sorbonne Universite, 4 place Jussieu, F-75005 Paris, France}


\begin{abstract}


Device-independent quantum key distribution (DIQKD) provides the strongest form of quantum security, as it allows two honest users to establish secure communication channels even when using fully uncharacterized quantum devices.
The security proof of DIQKD is derived from 
the violation of a Bell inequality, mitigating side-channel attacks by asserting the presence of nonlocality.
This enhanced security comes at the cost of a challenging implementation, especially over long distances, as losses make Bell tests difficult to conduct successfully.
Here, we propose a photonic realization of DIQKD, utilizing a heralded preparation of a single-photon path entangled state between the honest users.
Being based on single-photon interference effects, the obtained secret key rate scales with the square root of the quantum channel transmittance.
This leads to positive key rates over distances of up to hundreds of kilometers, making the proposed setup a promising candidate for securing long-distance communication in quantum networks. 

%
%

\end{abstract}
\maketitle


\section{Introduction}
Exchanging private communication in a network is a central feature of the modern world. 
Classical protocols are based on computational security, as secrecy  relies on 
computational assumptions.
Quantum key distribution (QKD)~\cite{Bennett84,Ekert91,Gisin2002,Lo2014,Scarani2009} provides quantum physical security, an alternative solution in which two parties measure quantum states and obtain correlated classical bits, from which a secure key is constructed.
No computational assumptions are needed, as these quantum correlations can be such that external adversaries cannot be correlated with the measurement outcomes, even when considering access to infinite computational power. 
The security of standard QKD protocols, however, relies on assumptions on the physics of the quantum devices used to distribute the key, in particular, it requires that these devices behave in the exact manner described by the protocol. 
In practice, verifying this assumption is challenging, as it demands accurate characterization of quantum states and measurements throughout the QKD protocol execution.  
In fact, by exploiting inaccurate quantum device calibrations, side-channel attacks have been successfully performed against QKD systems~\cite{Gerhardt2011,Lydersen2010,Zhao2008,Weier2011}.
To solve this critical problem, device-independent QKD (DIQKD) \cite{Mayers2004,Acin2007,Vazirani2014} protocols have been introduced: they use the violation of a Bell inequality to bound the information that adversaries may obtain on the distributed key, without requiring any modeling of the devices used in the protocol. DIQKD protocols therefore offer stronger security, as they are robust against any hacking attacks exploiting imperfections on the devices.

The implementation of DIQKD is challenging, as it requires the distribution of high-quality entangled states at long distances, as well as high-efficiency transmission channels and measurements. As photons in fibers are the natural carriers of quantum information, channel losses represent the main challenge for DIQKD: growing exponentially with distance, they become already at short distances too large for the honest users to be able to observe any Bell inequality violation~\cite{Acin2016}.
To circumvent channel losses, a heralding scheme can be used, where entanglement is generated between the parties' systems conditioned on the detection of photons at a central heralding station performing a joint measurement. Losses therefore reduce the key generation rate, but not its security. Heralding schemes have been used in recent proof-of-principle demonstrations of DIQKD~\cite{Nadlinger_2022,Zhang2022}. In these experiments, the honest users locally generate light-matter entanglement, so that the state at the local stations is encoded in material qubits. The photons are sent to the central station, where the joint measurement heralds the preparation of an entangled state between the local material qubits. These can be measured with nearly perfect efficiency, which allows a large Bell inequality violation. Despite these successful demonstrations, the light-matter entanglement generation process required in this approach typically has low repetition rates, limiting its scalability over large distances. Purely photonic platforms are arguably the most suited for high-rate long-distance DIQKD applications~\cite{Zapatero2023,Valcarce2023}. Several proposals for heralding photonic DIQKD exist, see for instance~\cite{Gisin10} or~\cite{Kolodynski2020,Gonzalez-Ruiz24}. However, these schemes require two-photon interference at the central heralding station, resulting in low repetition rates that make them impractical for long distances. 


In this article, we propose a photonic DIQKD  implementation that offers significant advantages with respect to the existing proposals. First of all, our scheme is based on single-photon interference effects. This results in much higher key rates, as they scale like the square root of the channel transmittance, as opposed to the previous schemes, which were based on two-photon interference and therefore had rates scaling like the channel transmittance. 
While this scaling can be found in other proposals \cite{Mycroft23,Xie2021}, our protocol presents important advantages that either simplify the implementation or offer stronger noise robustness. The proposal in \cite{Xie2021} relies on light-matter interaction, while ours is purely photonic. In comparison to \cite{Mycroft23}, we show an improvement of approximately 8$\%$ in the efficiencies required for the violation of the Clauser-Horne-Shimony-Holt (CHSH) Bell inequality.
Single-photon interference is also at the heart of twin field QKD \cite{Lucamarini2018}, a scheme that achieves the same scaling for the key rate but needs to assume fully characterized preparation devices, unlike our device-independent scenario. This is possible because in our protocol single-photon interference is used to distribute a single-photon entangled state between the honest users, which is then measured to obtain the Bell violation required for device-independent QKD. Our second main ingredient in fact consists of a new measurement scheme to observe Bell inequality violations from single-photon entangled states. This allows us to improve the robustness of the Bell test, in particular, the detection efficiencies needed for secure key distribution. 



The manuscript is organized as follows. In Sec.~\ref{sec:scenario}, we present the scenario, including the heralding scheme, the novel measurement strategy, and the physical model adopted for the single-photon sources. In Sec.~\ref{sec:protocol}, we describe the DIQKD protocol under consideration and the methods used in our simulations. In Sec.~\ref{sec:security}, we discuss the security of the protocol, both in the asymptotic and finite-size regime. In Sec.~\ref{sec:results}, we present the simulation results, focusing on the CHSH values achievable with our measurement scheme and key rates. In Sec. \ref{sec:feasibility}, we discuss the main challenges regarding the implementation of our protocol. Finally, in Sec.~\ref{sec:conclusions}, we conclude with a discussion of the advantages and limitations of the proposed approach.

\section{Scenario}\label{sec:scenario}
\subsection{Setup}
\begin{figure*}
    \centering
    \includegraphics[width=0.8\textwidth]{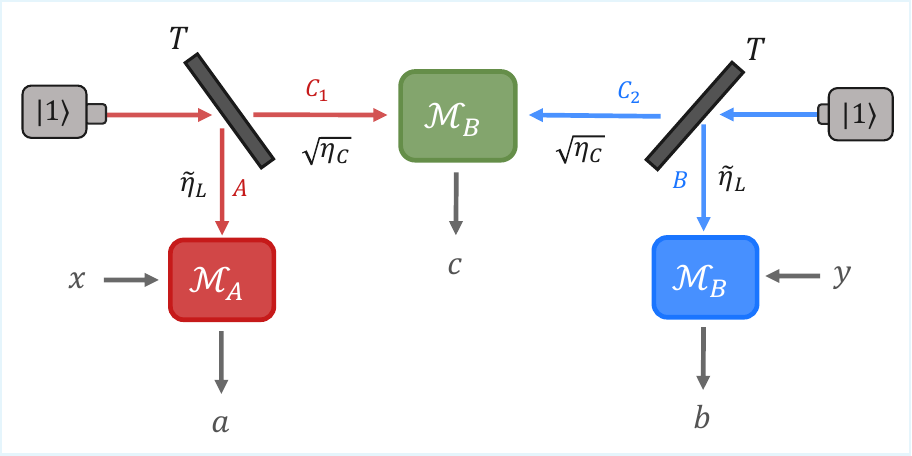}
    \caption{ Alice (in red) and Bob (in blue) possess a heralded single-photon source, where the generated photons are directed towards an unbalanced beamsplitter with low transmittance $T$. The transmitted photons travel through a symmetric lossy channel with efficiency $\sqrt{\eta}_C$ and arrive at a central station named Charlie, whose measurement system is denoted $\mathcal{M}_C$, with outcome $c$. The photons reflected by the unbalanced beamsplitter pass through another lossy channel characterised by $\tilde{\eta}_L$. All these losses are modeled as a beamsplitter where one of the modes is lost.  Detection events at Charlie's detectors herald the state in Eq. \eqref{eq: heralded state}. In this case, Alice and Bob start the DIQKD protocol according to the inputs $x$ and $y$ with measurement outcomes $a$ and $b$. } 
    \label{fig:setup}
\end{figure*}

We consider the scenario illustrated in Fig.~\ref{fig:setup}. Two honest users, Alice and Bob, possess a single-photon source that needs to be heralded, that is, such that Alice and Bob should know when the single photon has been produced. For the sake of clarity, we describe the setup assuming ideal single-photon sources. A realistic model for the sources used in the simulations will be introduced in Sec.~\ref{sec:realistisource}.
The emitted photons are directed towards a beamsplitter with transmittance $T$. This is a free parameter of the protocol that can be tuned to optimise the key rate, although in practice this transmittance needs to have a small value for reasons that will become clear below. Therefore, it is convenient for what follows to consider that $T$ is small. 

\nocite{Tan2022}
The transmitted modes are sent through a lossy channel to a measurement device $\mathcal{M}_C$, located at Charlie's. For simplicity, we assume that this device is placed at half the distance between Alice and Bob; therefore, we will also refer to Charlie as the central station.
Let us denote by $\eta_C$ the efficiency of the channel directly connecting Alice and Bob. The dependence of $\eta_C$ on the distance is 

    \begin{equation}
    \eta_C = 10^{-\alpha_{\text{att}} L/10},
    \end{equation}

\noindent where $L$ indicates the distance between Alice and Bob, and $\alpha_{\text{att}}$ denotes the attenuation coefficient of the medium through which the photons travel. For optical fibers at telecommunication wavelengths, $\alpha_{\text{att}}$ is typically $\SI{0.2}{\dB /\kilo\meter}$. Since the distance between each honest user and Charlie is $L/2$, the losses in the corresponding channels are the same and equal to $\sqrt{\eta_C}$. The reflected particle, A (B), is directly sent to the measurement systems of Alice (Bob). The bipartite state of Alice (Bob) and Charlie, expressed in the photon-number basis, is given by 
    
 \begin{align}
     \rho=\Big( \ketbra{\psi}{\psi}+T(1-\sqrt{\eta_C} )\ketbra{00}{00}\Big)_{AC_1/BC_2},
\end{align}

\noindent where $\ket{\psi} = \sqrt{T\sqrt{\eta_C}}\ket{01} + \sqrt{1-T}\ket{10}$. At Charlie's station, the modes $C_1$ and $C_2$ are combined into a balanced beamsplitter that erases the path information. The output modes are measured with photo-detectors of efficiency equal to $\eta_D$ (see Fig. \ref{Fig:charlie}). Charlie announces the measurement results, keeping those instances where only one of the two detectors clicks, while the others are discarded. When $T$ is small, a click on the left detector of Charlie heralds the state \footnote{Note that a detection on the right detector will herald the state $(\ket{10}-\ket{01})/\sqrt{2}$} 
    \begin{align}
        \ket{\Psi}_H = \frac{1}{\sqrt{2}}  (\ket{10} + \ket{01})_{AB}, \label{eq: heralded state}
    \end{align}
\noindent between Alice and Bob at leading order in $T$ (see \cite[Section B]{SM} for more details). This is because $T$ is chosen small enough so that the probability that one photon has been transmitted to Alice or Bob's locations and one detected by Charlie is much larger than the probability that the two photons are sent to Charlie and one is detected. At first order in $T$, a successful heralding happens with probability $P_H =T\eta_H$, with $\eta_H=\eta_D\sqrt{\eta_C}$. The entangled state~\eqref{eq: heralded state} is sent into the measurement devices $\mathcal{M}_A$ and $\mathcal{M}_B$ on Alice and Bob's locations to establish the secret key. This process is not ideal and adds some additional losses encapsulated by the local efficiency $\tilde{\eta}_L$. 

\subsection{Measurements} \label{sec:measurements}

As we detail below, the success of our scheme depends on how accurately we can perform Pauli measurements in the subspace spanned by the $\ket{0}$ and $\ket{1}$ photon Fock states. For measurements in the $z$-direction of the Bloch sphere, the measurement operators are the projectors $\{\ketbra{0}{0}, \ketbra{1}{1}\}$, which correspond, respectively, to no-click and click events of a single-photon detector. On the other hand, to perform measurements in superpositions of these basis states (i.e., the $x$–$y$ plane), we apply a parametric unitary transformation $U(\vec{s})$ before the single-photon detector, where $\vec{s}$ is a vector that parametrises the unitary. The resulting operators are defined as 
\begin{align}
    \Pi^{(\vec{s})}_{-1} &:=U(\vec{s})\ketbra{0}{0}U(\vec{s})^{\dagger}, \\
    \Pi^{(\vec{s})}_{+1} &:=\mathds{1} - (U(\vec{s})\ketbra{0}{0}U(\vec{s})^{\dagger}), 
\end{align}
again corresponding to no-click and click outcomes, respectively. When U is a displacement operator, one can project onto noisy superpositions of orthogonal photonic qubit states \cite{Banaszek_1999}. Furthermore, such measurements have been used to certify entanglement in setups similar to ours \cite{Caspar2020}. To quantify the fidelity of these imperfect measurements with respect to an ideal projective measurement, we can define the observable 
\begin{align} \label{eq: observable as a function of POVM}
\widetilde{\sigma}(\vec{s}) := \Pi^{(\vec{s})}_{+1} - \Pi^{(\vec{s})}_{-1}
\end{align}
\begin{figure*}
    \begin{minipage}{.5\textwidth}
        \subfloat[]{\includegraphics[width=\textwidth]{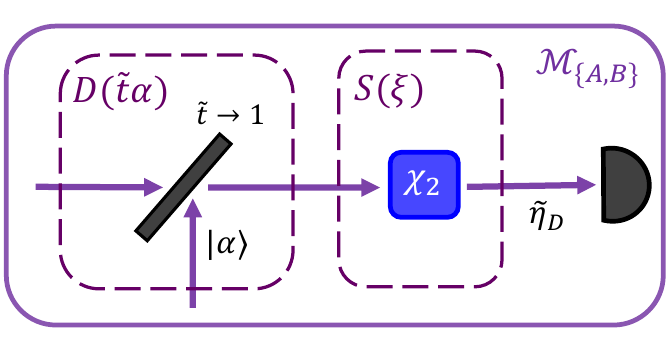} \label{Fig:measurements}}
    \end{minipage}
    \hfill    
    \begin{minipage}{.4\textwidth}
        \subfloat[]{\includegraphics[width=\textwidth]{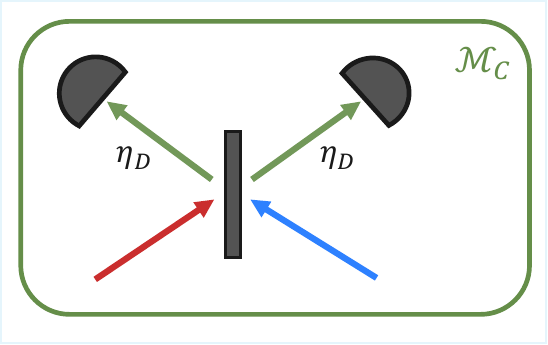} \label{Fig:charlie}}
    \end{minipage}
        \caption{(a) The heralded state \eqref{eq: heralded state} is sent to Alice's (Bob's) measurement system $\mathcal{M}_A$ ($\mathcal{M}_B$). The measurements comprise a displacement operation, implemented with a beamsplitter with transmittance $\tilde{t}$ close to one \cite{Paris1996DisplacementOB}, followed by a nonlinear crystal $\chi^{(2)}$ and a single-photon detector with efficiency $\tilde{\eta}_D$. The nonlinear crystal squeezes the incoming state, enhancing the distinguishability of an arbitrary qubit state spanned by 0 and 1 photons upon a click or no-click event at the detector. 
        (b) Charlie's measurement device, which consists of a $50/50$ beamsplitter and two single-photon detectors of efficiency $\eta_D$.}\label{fig:sub1}
    \end{figure*}
and introduce the overlap along a given direction in the Bloch sphere $R(\vec{s}) = \text{Tr}(\sigma_{x/y} \widetilde{\sigma}_{\text{Q}}(\vec{s}))$, where $\widetilde{\sigma}_{\text{Q}}(\vec{s})$ is the operator in \eqref{eq: observable as a function of POVM}, restricted to the qubit Fock subspace spanned by $\ket{0}$ and $\ket{1}$. The best achievable fidelity for a target Pauli measurement is denoted as $R_M = \max_s R(\vec{s})$. For displacement-only measurements, we find $R_M \approx 0.88$, a value that remains constant for all measurements in the $x$-$y$ plane of the Bloch sphere. To improve this result, we propose adding a squeezing operation after the displacement, which leads to a higher overlap $R_M \approx 0.94$. Therefore, for Alice’s and Bob’s measurements, we consider a setup consisting of an unbalanced beam splitter followed by a nonlinear crystal, implementing the displacement and squeezing operations, respectively. This is followed by a single-photon detector with efficiency $\tilde{\eta}_D$ (see Fig.~\ref{Fig:measurements}). As mentioned above, the squeezing operation increases the distinguishability between the relevant qubit superposition states (see also \cite[Section A]{SM}), which is crucial for reducing the detection efficiency required for our protocol to succeed.

\subsection{Realistic single-photon source}\label{sec:realistisource}

A crucial component of our protocol is the generation of high-quality single photons. We consider two types of sources: a heralded source using spontaneous parametric down-conversion (SPDC), where one photon of a generated pair is detected to signal the presence of the other, and a ``more intrinsic" single-photon source, such as those based on quantum dots (QDs). To characterise the quality of these sources, we introduce two key parameters. The first is the Hong-Ou-Mandel visibility $V$, which quantifies the indistinguishability of the produced photons and directly impacts interference visibility, thereby affecting the protocol’s performance. The second is the second-order correlation function $g^{(2)}$, which measures the probability of multi-photon emission events, with lower values indicating purer single-photon generation. In \cite[Section C]{SM}, we provide detailed models of both SPDC and QD sources, compare their characteristics through these parameters, and analyse how variations in $V$, $g^{(2)}$, and other experimental factors affect the system’s performance. This analysis supports the choice of QDs as the preferred source for our protocol, due to their superior performance over long distances.

\section{Device-Independent Quantum Key Distribution} \label{sec:protocol}
\subsection{Assumptions}

The security of device-independent protocols requires assumptions on the causal connections among devices, but is independent of their inner workings. Devices are, in fact, modeled as uncharacterized quantum black boxes processing classical inputs to produce classical outputs. In our setup, there are five devices: the state preparation and measuring devices of each party, and the measurement device at the central location. It is assumed that:
\begin{enumerate}
\item At round $i$, the measurement choices, denoted by $x_i$ for Alice and $y_i$ for Bob, are independent from any other variable generated in the protocol, for instance, because produced by proper (quantum) random number generators.
\item{At round $i$, the measurement outcome of each party, say $a_i$ by Alice, only depends on her measurement choice $x_i$, her prepared state at round $i$, and also on all the variables involved in the past measurements, that is, Alice's previous measurement choices and results, and prepared states in rounds $i'<i$. Similar assumptions apply to Bob's measurement outcomes.}
\item{The preparation of states, say by Alice, only depends on her previously prepared states. In particular, it is causally disconnected from her measurement processes.}
\item At the central location, the measurement process producing the output at round $i$ only depends on all the prepared states by Alice and Bob at rounds $i'\leq i$, and on all the previous measurement results, at rounds $i'<i$. 
\end{enumerate}
Note that it is assumed that any process taking place at Alice's location is disconnected from anything happening at Bob's location, and vice versa. Moreover, the different processes at each location, measurements and state preparations, depend on their respective pasts. 

Assumptions can never be proven to be true and their validity can of course always be questioned. The assumptions involving Alice and Bob's different locations seem quite plausible, given the physical distance and protocol configuration, while assumptions 1 and 3 may be more questionable. The honest users can also take measures to enforce their plausibility. In any case, in this work, these assumptions are taken as valid. The failure of any of them compromises the protocol security and, therefore, represents a security loophole.

We find convenient to illustrate the meaning of these assumptions in an iid, for independent and identically distributed, picture (which, however, is not assumed in the security proof below and is just discussed here for the sake of clarity). The corresponding distribution reads
\begin{equation}
p(a,b,c|x,y)=\tr\left[(M_{a|x}\otimes M_c\otimes M_{b|y})(\rho_{AC_1}\otimes\rho_{C_2B})\right],
\end{equation}
where $M_{a|x}$, $M_c$ and $M_{b|y}$ are the measurements at Alice, the central and Bob's locations, while $\rho_{A C_1}$ and $\rho_{C_2B}$ are the states prepared by Alice and Bob. Note that the process at the central location, consisting of measuring particles $C_1$ and $C_2$ prepared by Alice and Bob and subsequent post-selection of result $c=\bar c$, can simply be seen as the heralded preparation of the state, at Alice and Bob's locations,
\begin{equation}
\sigma_{AB}=\frac{1}{p(\bar c)}\tr_C\left[(\mathds{1}\otimes M_{\bar c}\otimes \mathds{1}) (\rho_{AC_1}\otimes\rho_{C_2B})\right] ,
\end{equation}
with probability $p(\bar c)=\tr\left[(\mathds{1}\otimes M_{\bar c}\otimes \mathds{1})(\rho_{AC_1}\otimes\rho_{C_2B})\right]$. In fact, the probabilities used to compute the CHSH inequality \cite{CHSH} and construct the key have the form
\begin{equation}
p(a,b|x,y,c=\bar c)=\tr\left[(M_{a|x}\otimes M_{b|y})\sigma_{AB}\right],
\end{equation}
as in any Bell test. When considering more rounds in a non-iid configuration, the validity of assumptions 1-4 ensure that the conditions needed to employ the existing security proofs for DIQKD are met in our setup.

\subsection{Protocol}
We consider the DIQKD protocol introduced in~\cite{Acin2007}, which is based on the violation of the CHSH inequality~\cite{CHSH}. This inequality has the form:
\begin{equation}
\label{eq:CHSH}
S = \langle A_1 B_1 \rangle + \langle A_1 B_2 \rangle + \langle A_2 B_1 \rangle - \langle A_2 B_2 \rangle \leq 2,
\end{equation}
where $A_x$ and $B_y$ denote the observables measured by Alice and Bob, respectively. The bound of 2 represents the maximum value attainable by local hidden variable models. The security of the protocol relies on the ability to bound the information that a quantum eavesdropper can have about Alice's outcomes, and thus about the secret key, from the observed CHSH violation. 

The protocol consists of four main phases, enumerated below:
\begin{enumerate}
    \item \emph{Acquisition phase}. Here, Alice and Bob perform $N$ sequential rounds. In each round, upon receiving a quantum state, they randomly select inputs $x\in\{0,1\}$, $y\in\{0,1,2\}$, and measure accordingly. The rounds are categorized into test rounds and key generation rounds.
    \begin{enumerate}
        \item Test rounds occur with probability $\gamma\in(0,1)$, and correspond to input pairs $x\in\{0,1\}$, $y\in\{0,1\}$. 
    In these rounds, Alice and Bob aim to maximize the CHSH violation \eqref{eq:CHSH}. During the parameter estimation phase, this value will be used to bound the information that an eavesdropper could have about the outcomes.
    
        \item Key generation rounds occur with probability $1-\gamma$, and correspond to inputs $x=0$ and $y=2$. These rounds are used to extract the raw key, and the goal is to maximize the correlation between Alice and Bob's outcomes, minimizing the error probability $p(a_i\neq b_i\vert x_i=0,y_i=2)$.
        Throughout all rounds, Alice and Bob privately store their inputs and outputs.
    \end{enumerate}
\item  \emph{Parameter estimation phase}. Once the $N$ rounds are complete, Alice and Bob publicly reveal their inputs and outputs for all test rounds and compute the observed CHSH value. If the value falls below a pre-agreed threshold, the protocol is aborted; otherwise, they proceed to error correction.
A small fraction of the key rounds is also announced, so that Alice and Bob can estimate the error rate. This parameter is critical for the error correction phase, while the value of the CHSH expression is used for privacy amplification.

\item \emph{Error correction phase}. Alice and Bob aim to reconcile their raw keys generated from the key generation rounds.
To this end, Alice computes and sends a classical error-correction syndrome $M$ to Bob, which allows him to identify and correct discrepancies between his outcomes and hers. After this phase, Alice and Bob share identical raw keys.

\item \emph{Privacy amplification phase}. Alice and Bob apply a classical randomness extractor to their shared keys to remove any partial knowledge that an eavesdropper may have gained during the previous steps. This produces a shorter, but provably secure, final secret key.
\end{enumerate}

\section{Security}
\label{sec:security}

To compute the expected key rate in the limit of an infinite number of protocol rounds, we use the Devetak-Winter bound~\cite{devetak2005distillation}

\begin{equation}
    r_\infty \geq H(A_1\vert E) - H(A_1\vert B_3),
\end{equation}

\noindent which provides a lower bound on the asymptotic key rate $r_\infty$ per key round of the DIQKD protocol. While it is possible to compute $H(A_1\vert B_3)$ directly from the statistics of the scenario previously described, $H(A_1\vert E)$ cannot be directly assessed.
Hence, we rely on the analytical lower bound derived in 
\cite{Noisy_Preprocessing_Ho20}, getting

\begin{align}
    \begin{split}
        r_\infty \geq & \,r_\text{DW} = 1 - h\left(\frac{1+\sqrt{(S/2)^2-1)}}{2}\right) - H(A_1\vert B_3) \\
        & + h\left(\frac{1+\sqrt{1-q_n(1-q_n)(8 - S^2)}}{2}\right),\label{eq:asymptotic}
    \end{split} 
\end{align}

\noindent where $h(X)$ represents the binary entropy of $X$, $S$ stands for the CHSH value \eqref{eq:CHSH}, and $q_n$ indicates the probability of Alice performing a bit-flip on her outcome. This noisy pre-processing step by Alice can indeed be used to optimize the lower bound on the key rate, as the maximum of \eqref{eq:asymptotic} is, in general, obtained for a non-zero value of $q_n$. Throughout the manuscript, we refer to the quantity defined in \eqref{eq:asymptotic} as the Devetak–Winter (DW) rate $r_\text{DW}$.

To evaluate the performance of our protocol, we simulate the Devetak–Winter rate under realistic conditions. In these simulations, we fix the detector efficiency and numerically optimize the remaining parameters to maximize $r_\text{DW}$ in Eq.~\eqref{eq:asymptotic}.
We first evaluated the rate as a function of the local transmission efficiency $\tilde{\eta}_L$ at zero distance ($L=$ \SI{0}{\kilo\meter}), numerically optimizing Eq.\eqref{eq:asymptotic} using the measurement settings $\mathcal{M}_A$ and $\mathcal{M}_B$ shown in Fig.\ref{Fig:measurements} and detailed in \cite[Section A]{SM}. Then, keeping $\tilde{\eta}_L$ fixed, we evaluated the final key rate per unit of time as a function of distance $L$ by multiplying $r_\text{DW}$ by the heralding rate $R_H$ and the source’s generation rate $\nu$, and repeating the numerical optimization for each distance.

In practical implementations, the number of rounds $N$ is finite, which introduces finite-size effects. These reduce the length of the final secure key compared to the asymptotic case. To account for this, we apply the entropy accumulation theorem (EAT) \cite{Dupuis_2020}, following the approach of \cite{Nadlinger_2022}, which bounds the eavesdropper’s knowledge by computing a per-round entropy contribution and applying a correction term that depends on $N$.

The key parameters in this analysis are the soundness and completeness parameters, $\varepsilon_\text{sound}$ and $\varepsilon_\text{com}$~\cite{Tan2022}. The soundness parameter bounds the distance between the actual and ideal key states, while the completeness parameter bounds the probability of aborting when the protocol is executed with honest devices. This parameter ensures that the protocol does not excessively terminate or abort when executed under normal, honest conditions. For our analysis we fixed these parameters to  $\varepsilon_\text{sound}=3 \cdot 10^{-10}$ and $\varepsilon_\text{com}<10^{-2}$.
The resulting number of secret bits per round is defined as $r_\text{FS} = \ell / N$, where $\ell$ is the length of the final secure key. As $N \to \infty$, $r_\text{FS}$ converges to $r_\text{DW}$. A detailed description of the relevant parameters, functions and corrections used to compute $\ell$ is provided in \cite[Section E]{SM}.
The final key rate is given by
\begin{equation}
R_\text{FS} = R_H \, r_\text{FS} \, \nu, \label{eq: finite key rate}
\end{equation}
where in our protocol, the heralding rate is $R_H=\, T \, (1-T) \, \eta_H$. Note that the source's generation rate depends on the choice of single-photon source.

\section{Results}\label{sec:results}

\subsection{CHSH value}
\begin{figure}
    \centering
    \includegraphics[width=0.5\textwidth]{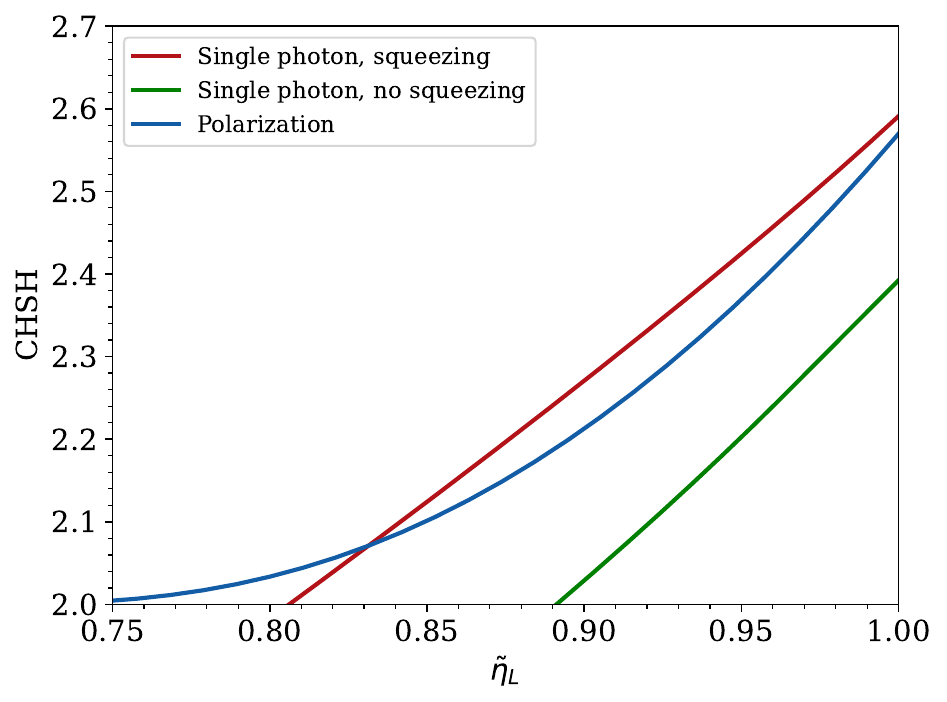}
    \caption{Comparison of CHSH values as a function of $\tilde{\eta}_L$, for a fixed detector efficiency $\tilde{\eta}_D = 95\%$. For our measurement scheme with squeezing (red), the channel threshold is $\tilde{\eta}_L = 80.7\%$, corresponding to an overall local efficiency threshold of $\eta_L = \tilde{\eta}_L \tilde{\eta}_D = 76.7\%$. Without squeezing (green), the threshold is $\tilde{\eta}_L = 89.2\%$, or $\eta_L = 84.7\%$.} 
    \label{fig:CHSH_comparison}
\end{figure}

To benchmark the performance of our measurements $\mathcal{M}_A$ and $\mathcal{M}_B$, described in Section \ref{sec:measurements} and shown in Fig. \ref{fig:sub1}, we numerically computed the maximum CHSH achievable with them when considering the ideal heralded state \eqref{eq: heralded state},  subject to local noise modeled as $\rho = \tilde{\eta}_L \ketbra{\Psi}{\Psi}_H + (1 - \tilde{\eta}_L) \ketbra{00}{00}$. Specifically, we considered realistic measurement implementations as described in \cite[Section A]{SM}, both with and without the inclusion of a squeezing operation, in addition to displacement. For comparison, we also evaluated the CHSH value achievable in a polarization-encoded scenario with the state $\ket{\psi} = \beta \ket{HV} + \sqrt{1 - \beta^2} \ket{VH}$, optimizing over the free parameter $\beta$. Local noise in the polarization case was modeled as in \cite{Kolodynski2020}.
For all these cases, the detector efficiency is fixed to $95\%$. The result of this simulation is reported in Fig.~\ref{fig:CHSH_comparison}.

We used a maximum squeezing amplitude of \SI{2.95}{dB} and a displacement amplitude of $0.6$, both achievable with current experimental technology. Notably, the green curve reproduces the violation obtained without squeezing, as in \cite{Mycroft23}. The inclusion of squeezing consistently yields significantly higher CHSH values and improved robustness efficiency compared to the measurement proposed in \cite{Mycroft23}, across all values of $\tilde{\eta}_L$. Although polarization encoding is more robust to noise, our squeezed scheme surpasses it in CHSH performance for $\tilde{\eta}_L \gtrsim 83\%$, which corresponds to the regime where a positive key rate becomes attainable. Therefore, our analysis shows that squeezing is not merely an enhancement but a necessary resource to outperform polarization-based schemes. To the best of our knowledge, this constitutes the highest CHSH violation reported for efficiencies exceeding 83\%.

\subsection{Key rates}
For analyzing the key rates, we considered both realistic measurements and realistic photon sources. In particular, as detailed in \cite[Section C]{SM}, we consider QDs. For the analysis we consider sources with second-order correlation function $g^{(2)}=0.01$, Hong-Ou-Mandel visibility $V=97.5\%$ and a generation rate of \SI{75}{\mega\hertz}~\cite{Tomm_2021,Uppu20,Wang19}. Concerning the measurements, we consider detectors with efficiency $95\%$ and a dark-count rate on the central heralding station of 1 count/s. Finally, we consider a transmittance $T=0.01$. The motivation behind this choice of parameters is detailed in \cite[Section C]{SM}. From the asymptotic analysis using Eq.~\eqref{eq:asymptotic}, we found that the minimum total local efficiency required to achieve a positive DW rate is $\eta_L = \tilde{\eta}_L \tilde{\eta}_D = 90.4\%$.

\begin{figure}
    \centering
    \includegraphics[width=0.5\textwidth]{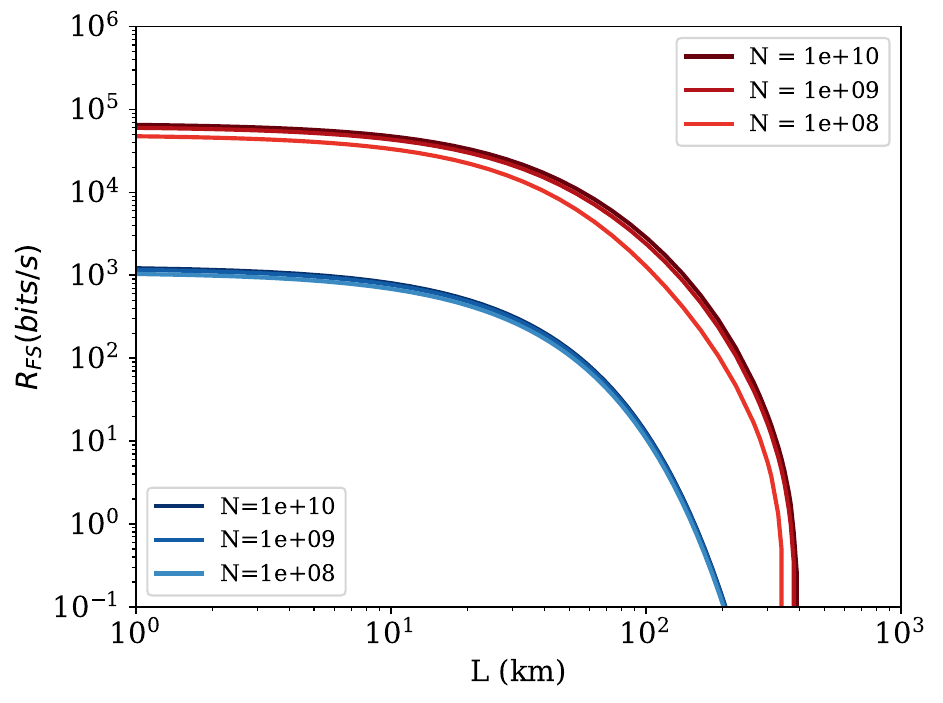}
    \caption{Key rate as a function of the distance $L$ for various numbers of rounds $N$, using the DIQKD protocol proposed in this work (red) and the polarization-based scheme from~\cite{Kolodynski2020,Gonzalez-Ruiz24} (blue). In all calculations we fixed $\tilde{\eta}_D = 95\%$, $\tilde{\eta}_L = 98.3\%$ (i.e., $\eta_L = 93.4\%$), $T = 0.01$,
    $\nu=$\SI{75}{\mega\hertz} and $\epsilon_\text{sound} = 3 \cdot 10^{-10}$. Our protocol includes realistic imperfections such as non-ideal single-photon sources with $g^{(2)} = 0.01$ and $V = 97.5\%$, and dark-count corrections (1 dark-count/s for the detector at the heralding station). In contrast, the polarization scheme is modeled using an ideal entangled state, neglecting imperfections other than local losses and channel attenuation.}
    \label{fig:key_vs_L}
\end{figure}

To further evaluate the performance of our protocol in a realistic setting, we studied the finite-size key rate as a function of distance for various numbers of rounds $N$. For this analysis, we fixed $\tilde{\eta}_L = 98.3\%$, yielding a total local efficiency of $\eta_L = \tilde{\eta}_D \tilde{\eta}_L = 93.4\%$. This choice ensures that the DW rate satisfies $r_\text{DW} > 10^{-1}$.
We compare the performance of our protocol with that of the polarization-based scheme proposed in~\cite{Gonzalez-Ruiz24}. While our simulation includes realistic imperfections such as source non-idealities and dark counts, the polarization-based scheme is analyzed in a more idealized setting: the only imperfections considered are local losses (with the same $\eta_L = 93.4\%$) and channel transmittance, whereas other sources of noise are neglected.
For both simulations, we imposed a soundness error of $\epsilon_\text{sound} = 3 \cdot 10^{-10}$. The results of this comparison are shown in Fig.~\ref{fig:key_vs_L}.

We observe that, even when the polarization-based protocol is considered under idealized assumptions (blue lines), our scheme (red lines) yields significantly higher key rates and enables communication over longer distances. This improvement stems from the fact that polarization-based schemes require two-photon interference at the central station, leading to a heralding rate that scales as
\begin{equation}
    R_H = T^2 \, \eta_C \, \eta_D^2 \;\nu, \label{eq: heralding probability one-photon case}
\end{equation}
\noindent i.e., proportional to the channel transmittance $\eta_C$.
In contrast, our single-photon heralding scheme scales more favorably with $\sqrt{\eta_C}$, leading to a substantial gain in performance. Specifically, using our schemes, key rates of the order of kb/s are obtained for distances of the order of $\SI{100}{\kilo\meter}$ and for block sizes of $10^8$, which become unattainable already at short distances for the polarisation-based protocol. This demonstrates the practical advantage of our approach for long-distance device-independent quantum key distribution.

\section{Feasibility} \label{sec:feasibility}
Although this protocol involves numerous technically demanding tasks, recent experimental progress provides a promising pathway toward the feasibility of our scheme. More specifically, our proposal relies on two main challenging ingredients: the successful distribution of single-photon entanglement between two distant locations and the implementation of local measurements based on displacement and squeezing, followed by a single-photon detector (see Fig. \ref{Fig:measurements}). Regarding the first, a significant challenge is the phase-locking of the two signals arriving at the central heralding station. This issue has been addressed in \cite{Lago-Rivera2021}, where a two-photon interference heralding scheme is utilized. The authors inject a classical reference field and make it interfere with the heralding signal during the so-called \emph{locking rounds}. By measuring this interference, they implement a feedback loop to correct for any phase drift effectively. Similarly, \cite{Caspar2020} demonstrated phase stabilization using a single-photon interference heralding scheme. Furthermore, single-photon heralding schemes have been successfully implemented to achieve solid-state entanglement at a metropolitan scale, not only with various phase-locking solutions but also compensating for polarisation and fiber length drift \cite{stolk2021,Liu2024}. Recent large-scale experiments also demonstrate the viability of phase stabilization over ultra-long distances. In the context of twin field QKD, the experiment reported in \cite{1000kmTwinField} showcases phase-locking using two laser wavelengths, one for the signal and another for phase reference, without the need for active modulators or real-time feedback.  
Regarding the second challenge, several experiments have demonstrated single-photon entanglement using measurements consisting of displacement, as described in the caption of Fig. \ref{Fig:measurements}, and photo-detection \cite{Caspar2020, Caspar2022, Monteiro}. To implement our measurements, it is required to add a squeezing operation, which seems possible, taking into account that the demanded level of squeezing is of the order of \SI{2.95}{dB}, within reach of current setups \cite{Andersen30YearsSqueezing}. Note that the experiment proposed in \cite{Caspar2020} is quite close to our proposal when using SPDC sources, as it reports the observation of heralded single-photon entanglement between two locations separated by a distance of \SI{2}{\kilo\meter} using measurements consisting of displacement and photo-detectors.

\section{Conclusions}\label{sec:conclusions}
This work presents a practical proposal for photonic DIQKD that significantly advances the state of the art. By combining the favourable rate-distance scaling of twin-field QKD with the security guarantees of DIQKD, our protocol achieves key rates that scale with the square root of the transmittance between the honest users, while not requiring any characterisation of the devices. A key ingredient behind our results is the proposed local measurement scheme, comprised of a displacement and squeezing operator, followed by a single-photon detector. Using maximum displacement amplitudes of 0.6 and for squeezing up to 2.95 dB, both experimentally achievable, we observe higher CHSH values and improved robustness efficiencies compared to displacement-only measurements. We also compared our results with a polarisation encoding setup, finding that our scheme surpasses it in CHSH values in the regime where a positive key rate can be attained. Although our protocol is source-independent, simulations require specifying a concrete source model. We based our simulations using typical parameters of quantum dot sources.
In this context, we performed a finite-size analysis of the key rates, fixing local efficiencies in Alice's and Bob's laboratories to 93.4\% in order to obtain key rates above $10^{-1}$ bits/s. In particular, we obtain key rates of the order of kb/s for distances around 100 km with a block size of $10^8$, which for polarisation-based protocols become unachievable already at short distances under similar conditions. Putting all these considerations together, our results provide a robust and experimentally grounded path for achieving photonic secure communications over long distances through DIQKD.

\section*{Acknowledgements}
This work is supported by the Government of Spain (Severo Ochoa CEX2019-000910-S, FUNQIP, NextGenerationEU PRTR-C17.I1), EU projects QSNP, Quantera Veriqtas, Fundació Cellex, Fundació Mir-Puig, Generalitat de Catalunya (CERCA program), the ERC AdG CERQUTE and the AXA Chair in Quantum Information Science. A.S. acknowledges support from the “Agencia Estatal de Investigación” (Ref. PRE2022-101475). M.N. acknowledges funding from the European Union’s Horizon Europe research and innovation programme under the MSCA Grant Agreement No. 101081441. X.V. acknowledges funding by the European Union’s Horizon Europe research and innovation program under the project “Quantum Security Networks Partnership” (QSNP, Grant Agreement No. 101114043). X.V and E.O acknowledge funding by the French national quantum initiative managed by Agence Nationale de la Recherche in the framework of France 2030 with the reference ANR-22-PETQ-0009. 

\section*{Code availability}
Code is available via Github at ref. \cite{git}.


\bibliography{note}

\begin{thebibliography}{13}%
\makeatletter
\providecommand \@ifxundefined [1]{%
 \@ifx{#1\undefined}
}%
\providecommand \@ifnum [1]{%
 \ifnum #1\expandafter \@firstoftwo
 \else \expandafter \@secondoftwo
 \fi
}%
\providecommand \@ifx [1]{%
 \ifx #1\expandafter \@firstoftwo
 \else \expandafter \@secondoftwo
 \fi
}%
\providecommand \natexlab [1]{#1}%
\providecommand \enquote  [1]{``#1''}%
\providecommand \bibnamefont  [1]{#1}%
\providecommand \bibfnamefont [1]{#1}%
\providecommand \citenamefont [1]{#1}%
\providecommand \href@noop [0]{\@secondoftwo}%
\providecommand \href [0]{\begingroup \@sanitize@url \@href}%
\providecommand \@href[1]{\@@startlink{#1}\@@href}%
\providecommand \@@href[1]{\endgroup#1\@@endlink}%
\providecommand \@sanitize@url [0]{\catcode `\\12\catcode `\$12\catcode `\&12\catcode `\#12\catcode `\^12\catcode `\_12\catcode `\%12\relax}%
\providecommand \@@startlink[1]{}%
\providecommand \@@endlink[0]{}%
\providecommand \url  [0]{\begingroup\@sanitize@url \@url }%
\providecommand \@url [1]{\endgroup\@href {#1}{\urlprefix }}%
\providecommand \urlprefix  [0]{URL }%
\providecommand \Eprint [0]{\href }%
\providecommand \doibase [0]{https://doi.org/}%
\providecommand \selectlanguage [0]{\@gobble}%
\providecommand \bibinfo  [0]{\@secondoftwo}%
\providecommand \bibfield  [0]{\@secondoftwo}%
\providecommand \translation [1]{[#1]}%
\providecommand \BibitemOpen [0]{}%
\providecommand \bibitemStop [0]{}%
\providecommand \bibitemNoStop [0]{.\EOS\space}%
\providecommand \EOS [0]{\spacefactor3000\relax}%
\providecommand \BibitemShut  [1]{\csname bibitem#1\endcsname}%
\let\auto@bib@innerbib\@empty
\bibitem [{\citenamefont {Vivoli}\ \emph {et~al.}(2015)\citenamefont {Vivoli}, \citenamefont {Sekatski}, \citenamefont {Bancal}, \citenamefont {Lim}, \citenamefont {Martin}, \citenamefont {Thew}, \citenamefont {Hugo.Zbinden}, \citenamefont {Nicolas.Gisin},\ and\ \citenamefont {Nicola.Sangouard}}]{CapraraVivoli2015}%
  \BibitemOpen
  \bibfield  {author} {\bibinfo {author} {\bibfnamefont {V.~C.}\ \bibnamefont {Vivoli}}, \bibinfo {author} {\bibfnamefont {P.}~\bibnamefont {Sekatski}}, \bibinfo {author} {\bibfnamefont {J.}~\bibnamefont {Bancal}}, \bibinfo {author} {\bibfnamefont {C.~W.}\ \bibnamefont {Lim}}, \bibinfo {author} {\bibfnamefont {A.}~\bibnamefont {Martin}}, \bibinfo {author} {\bibfnamefont {R.}~\bibnamefont {Thew}}, \bibinfo {author} {\bibnamefont {Hugo.Zbinden}}, \bibinfo {author} {\bibnamefont {Nicolas.Gisin}},\ and\ \bibinfo {author} {\bibnamefont {Nicola.Sangouard}},\ }\bibfield  {title} {\bibinfo {title} {Comparing different approaches for generating random numbers device‐independently using a photon pair source},\ }\href {https://doi.org/10.1088/1367-2630/17/2/023023} {\bibfield  {journal} {\bibinfo  {journal} {New Journal of Physics}\ }\textbf {\bibinfo {volume} {17}},\ \bibinfo {pages} {023023} (\bibinfo {year} {2015})}\BibitemShut {NoStop}%
\bibitem [{\citenamefont {Steffinlongo}(2024)}]{git}%
  \BibitemOpen
  \bibfield  {author} {\bibinfo {author} {\bibfnamefont {A.}~\bibnamefont {Steffinlongo}},\ }\href {https://github.com/AnnaSteffinlongo/DIQKD-with-single-photons} {\bibinfo {title} {https://github.com/annasteffinlongo/diqkd-with-single-photons}} (\bibinfo {year} {2024})\BibitemShut {NoStop}%
\bibitem [{\citenamefont {Liu}\ \emph {et~al.}(2022)\citenamefont {Liu}, \citenamefont {Zhang}, \citenamefont {Zhen}, \citenamefont {Li}, \citenamefont {Liu}, \citenamefont {Fan}, \citenamefont {Xu}, \citenamefont {Zhang},\ and\ \citenamefont {Pan}}]{SPDC}%
  \BibitemOpen
  \bibfield  {author} {\bibinfo {author} {\bibfnamefont {W.-Z.}\ \bibnamefont {Liu}}, \bibinfo {author} {\bibfnamefont {Y.-Z.}\ \bibnamefont {Zhang}}, \bibinfo {author} {\bibfnamefont {Y.-Z.}\ \bibnamefont {Zhen}}, \bibinfo {author} {\bibfnamefont {M.-H.}\ \bibnamefont {Li}}, \bibinfo {author} {\bibfnamefont {Y.}~\bibnamefont {Liu}}, \bibinfo {author} {\bibfnamefont {J.}~\bibnamefont {Fan}}, \bibinfo {author} {\bibfnamefont {F.}~\bibnamefont {Xu}}, \bibinfo {author} {\bibfnamefont {Q.}~\bibnamefont {Zhang}},\ and\ \bibinfo {author} {\bibfnamefont {J.-W.}\ \bibnamefont {Pan}},\ }\bibfield  {title} {\bibinfo {title} {Toward a photonic demonstration of device-independent quantum key distribution},\ }\href {https://doi.org/10.1103/PhysRevLett.129.050502} {\bibfield  {journal} {\bibinfo  {journal} {Phys. Rev. Lett.}\ }\textbf {\bibinfo {volume} {129}},\ \bibinfo {pages} {050502} (\bibinfo {year} {2022})}\BibitemShut {NoStop}%
\bibitem [{\citenamefont {Tomm}\ \emph {et~al.}(2021)\citenamefont {Tomm}, \citenamefont {Javadi}, \citenamefont {Antoniadis}, \citenamefont {Najer}, \citenamefont {Löbl}, \citenamefont {Korsch}, \citenamefont {Schott}, \citenamefont {Valentin}, \citenamefont {Wieck}, \citenamefont {Ludwig},\ and\ \citenamefont {Warburton}}]{Tomm_2021}%
  \BibitemOpen
  \bibfield  {author} {\bibinfo {author} {\bibfnamefont {N.}~\bibnamefont {Tomm}}, \bibinfo {author} {\bibfnamefont {A.}~\bibnamefont {Javadi}}, \bibinfo {author} {\bibfnamefont {N.~O.}\ \bibnamefont {Antoniadis}}, \bibinfo {author} {\bibfnamefont {D.}~\bibnamefont {Najer}}, \bibinfo {author} {\bibfnamefont {M.~C.}\ \bibnamefont {Löbl}}, \bibinfo {author} {\bibfnamefont {A.~R.}\ \bibnamefont {Korsch}}, \bibinfo {author} {\bibfnamefont {R.}~\bibnamefont {Schott}}, \bibinfo {author} {\bibfnamefont {S.~R.}\ \bibnamefont {Valentin}}, \bibinfo {author} {\bibfnamefont {A.~D.}\ \bibnamefont {Wieck}}, \bibinfo {author} {\bibfnamefont {A.}~\bibnamefont {Ludwig}},\ and\ \bibinfo {author} {\bibfnamefont {R.~J.}\ \bibnamefont {Warburton}},\ }\bibfield  {title} {\bibinfo {title} {A bright and fast source of coherent single photons},\ }\href {https://doi.org/10.1038/s41565-020-00831-x} {\bibfield  {journal} {\bibinfo  {journal} {Nature Nanotechnology}\ }\textbf {\bibinfo {volume} {16}},\ \bibinfo {pages} {399–403}
  (\bibinfo {year} {2021})}\BibitemShut {NoStop}%
\bibitem [{\citenamefont {Gonz\'alez-Ruiz}\ \emph {et~al.}(2022)\citenamefont {Gonz\'alez-Ruiz}, \citenamefont {Das}, \citenamefont {Lodahl},\ and\ \citenamefont {S\o{}rensen}}]{Gonzalez2022}%
  \BibitemOpen
  \bibfield  {author} {\bibinfo {author} {\bibfnamefont {E.~M.}\ \bibnamefont {Gonz\'alez-Ruiz}}, \bibinfo {author} {\bibfnamefont {S.~K.}\ \bibnamefont {Das}}, \bibinfo {author} {\bibfnamefont {P.}~\bibnamefont {Lodahl}},\ and\ \bibinfo {author} {\bibfnamefont {A.~S.}\ \bibnamefont {S\o{}rensen}},\ }\bibfield  {title} {\bibinfo {title} {Violation of bell's inequality with quantum-dot single-photon sources},\ }\href {https://doi.org/10.1103/PhysRevA.106.012222} {\bibfield  {journal} {\bibinfo  {journal} {Phys. Rev. A}\ }\textbf {\bibinfo {volume} {106}},\ \bibinfo {pages} {012222} (\bibinfo {year} {2022})}\BibitemShut {NoStop}%
\bibitem [{\citenamefont {Christ}\ \emph {et~al.}(2011)\citenamefont {Christ}, \citenamefont {Brecht}, \citenamefont {Mauerer},\ and\ \citenamefont {Silberhorn}}]{Christ2011}%
  \BibitemOpen
  \bibfield  {author} {\bibinfo {author} {\bibfnamefont {A.}~\bibnamefont {Christ}}, \bibinfo {author} {\bibfnamefont {B.}~\bibnamefont {Brecht}}, \bibinfo {author} {\bibfnamefont {W.}~\bibnamefont {Mauerer}},\ and\ \bibinfo {author} {\bibfnamefont {C.}~\bibnamefont {Silberhorn}},\ }\bibfield  {title} {\bibinfo {title} {Optimized generation of heralded fock states using parametric down-conversion},\ }\href {https://doi.org/10.1088/1367-2630/13/3/033027} {\bibfield  {journal} {\bibinfo  {journal} {New Journal of Physics}\ }\textbf {\bibinfo {volume} {13}},\ \bibinfo {pages} {033027} (\bibinfo {year} {2011})}\BibitemShut {NoStop}%
\bibitem [{\citenamefont {Muljarov}\ and\ \citenamefont {Zimmermann}(2004)}]{Muljarov2004}%
  \BibitemOpen
  \bibfield  {author} {\bibinfo {author} {\bibfnamefont {E.~A.}\ \bibnamefont {Muljarov}}\ and\ \bibinfo {author} {\bibfnamefont {R.}~\bibnamefont {Zimmermann}},\ }\bibfield  {title} {\bibinfo {title} {Dephasing in quantum dots: Quadratic coupling to acoustic phonons},\ }\href {https://doi.org/10.1103/PhysRevLett.93.237401} {\bibfield  {journal} {\bibinfo  {journal} {Phys. Rev. Lett.}\ }\textbf {\bibinfo {volume} {93}},\ \bibinfo {pages} {237401} (\bibinfo {year} {2004})}\BibitemShut {NoStop}%
\bibitem [{\citenamefont {Uppu}\ \emph {et~al.}(2020)\citenamefont {Uppu}, \citenamefont {Pedersen}, \citenamefont {Wang}, \citenamefont {Olesen}, \citenamefont {Papon}, \citenamefont {Zhou}, \citenamefont {Midolo}, \citenamefont {Scholz}, \citenamefont {Wieck}, \citenamefont {Ludwig},\ and\ \citenamefont {Lodahl}}]{Uppu20}%
  \BibitemOpen
  \bibfield  {author} {\bibinfo {author} {\bibfnamefont {R.}~\bibnamefont {Uppu}}, \bibinfo {author} {\bibfnamefont {F.~T.}\ \bibnamefont {Pedersen}}, \bibinfo {author} {\bibfnamefont {Y.}~\bibnamefont {Wang}}, \bibinfo {author} {\bibfnamefont {C.~T.}\ \bibnamefont {Olesen}}, \bibinfo {author} {\bibfnamefont {C.}~\bibnamefont {Papon}}, \bibinfo {author} {\bibfnamefont {X.}~\bibnamefont {Zhou}}, \bibinfo {author} {\bibfnamefont {L.}~\bibnamefont {Midolo}}, \bibinfo {author} {\bibfnamefont {S.}~\bibnamefont {Scholz}}, \bibinfo {author} {\bibfnamefont {A.~D.}\ \bibnamefont {Wieck}}, \bibinfo {author} {\bibfnamefont {A.}~\bibnamefont {Ludwig}},\ and\ \bibinfo {author} {\bibfnamefont {P.}~\bibnamefont {Lodahl}},\ }\bibfield  {title} {\bibinfo {title} {Scalable integrated single-photon source},\ }\href {https://doi.org/10.1126/sciadv.abc8268} {\bibfield  {journal} {\bibinfo  {journal} {Science Advances}\ }\textbf {\bibinfo {volume} {6}},\ \bibinfo {pages} {eabc8268} (\bibinfo {year} {2020})}\BibitemShut {NoStop}%
\bibitem [{\citenamefont {Zhao}\ \emph {et~al.}(2020)\citenamefont {Zhao}, \citenamefont {Ma}, \citenamefont {R\"using},\ and\ \citenamefont {Mookherjea}}]{Zhao2020}%
  \BibitemOpen
  \bibfield  {author} {\bibinfo {author} {\bibfnamefont {J.}~\bibnamefont {Zhao}}, \bibinfo {author} {\bibfnamefont {C.}~\bibnamefont {Ma}}, \bibinfo {author} {\bibfnamefont {M.}~\bibnamefont {R\"using}},\ and\ \bibinfo {author} {\bibfnamefont {S.}~\bibnamefont {Mookherjea}},\ }\bibfield  {title} {\bibinfo {title} {High quality entangled photon pair generation in periodically poled thin-film lithium niobate waveguides},\ }\href {https://doi.org/10.1103/PhysRevLett.124.163603} {\bibfield  {journal} {\bibinfo  {journal} {Phys. Rev. Lett.}\ }\textbf {\bibinfo {volume} {124}},\ \bibinfo {pages} {163603} (\bibinfo {year} {2020})}\BibitemShut {NoStop}%
\bibitem [{\citenamefont {Wang}\ \emph {et~al.}(2019)\citenamefont {Wang}, \citenamefont {He}, \citenamefont {Chung}, \citenamefont {Hu}, \citenamefont {Yu}, \citenamefont {Chen}, \citenamefont {Ding}, \citenamefont {Chen}, \citenamefont {Qin}, \citenamefont {Yang}, \citenamefont {Liu}, \citenamefont {Duan}, \citenamefont {Li}, \citenamefont {Gerhardt}, \citenamefont {Winkler}, \citenamefont {Jurkat}, \citenamefont {Wang}, \citenamefont {Gregersen}, \citenamefont {Huo}, \citenamefont {Dai}, \citenamefont {Yu}, \citenamefont {H{\"o}fling}, \citenamefont {Lu},\ and\ \citenamefont {Pan}}]{Wang19}%
  \BibitemOpen
  \bibfield  {author} {\bibinfo {author} {\bibfnamefont {H.}~\bibnamefont {Wang}}, \bibinfo {author} {\bibfnamefont {Y.-M.}\ \bibnamefont {He}}, \bibinfo {author} {\bibfnamefont {T.}~\bibnamefont {Chung}}, \bibinfo {author} {\bibfnamefont {H.}~\bibnamefont {Hu}}, \bibinfo {author} {\bibfnamefont {Y.}~\bibnamefont {Yu}}, \bibinfo {author} {\bibfnamefont {S.}~\bibnamefont {Chen}}, \bibinfo {author} {\bibfnamefont {X.}~\bibnamefont {Ding}}, \bibinfo {author} {\bibfnamefont {M.-C.}\ \bibnamefont {Chen}}, \bibinfo {author} {\bibfnamefont {J.}~\bibnamefont {Qin}}, \bibinfo {author} {\bibfnamefont {X.}~\bibnamefont {Yang}}, \bibinfo {author} {\bibfnamefont {R.-Z.}\ \bibnamefont {Liu}}, \bibinfo {author} {\bibfnamefont {Z.-C.}\ \bibnamefont {Duan}}, \bibinfo {author} {\bibfnamefont {J.-P.}\ \bibnamefont {Li}}, \bibinfo {author} {\bibfnamefont {S.}~\bibnamefont {Gerhardt}}, \bibinfo {author} {\bibfnamefont {K.}~\bibnamefont {Winkler}}, \bibinfo {author} {\bibfnamefont {J.}~\bibnamefont {Jurkat}}, \bibinfo {author}
  {\bibfnamefont {L.-J.}\ \bibnamefont {Wang}}, \bibinfo {author} {\bibfnamefont {N.}~\bibnamefont {Gregersen}}, \bibinfo {author} {\bibfnamefont {Y.-H.}\ \bibnamefont {Huo}}, \bibinfo {author} {\bibfnamefont {Q.}~\bibnamefont {Dai}}, \bibinfo {author} {\bibfnamefont {S.}~\bibnamefont {Yu}}, \bibinfo {author} {\bibfnamefont {S.}~\bibnamefont {H{\"o}fling}}, \bibinfo {author} {\bibfnamefont {C.-Y.}\ \bibnamefont {Lu}},\ and\ \bibinfo {author} {\bibfnamefont {J.-W.}\ \bibnamefont {Pan}},\ }\bibfield  {title} {\bibinfo {title} {Towards optimal single-photon sources from polarized microcavities},\ }\href {https://doi.org/10.1038/s41566-019-0494-3} {\bibfield  {journal} {\bibinfo  {journal} {Nature Photonics}\ }\textbf {\bibinfo {volume} {13}},\ \bibinfo {pages} {770–775} (\bibinfo {year} {2019})}\BibitemShut {NoStop}%
\bibitem [{\citenamefont {Nadlinger}\ \emph {et~al.}(2022)\citenamefont {Nadlinger}, \citenamefont {Drmota}, \citenamefont {Nichol}, \citenamefont {Araneda}, \citenamefont {Main}, \citenamefont {Srinivas}, \citenamefont {Lucas}, \citenamefont {Ballance}, \citenamefont {Ivanov}, \citenamefont {Tan}, \citenamefont {Sekatski}, \citenamefont {Urbanke}, \citenamefont {Renner}, \citenamefont {Sangouard},\ and\ \citenamefont {Bancal}}]{Nadlinger_2022}%
  \BibitemOpen
  \bibfield  {author} {\bibinfo {author} {\bibfnamefont {D.~P.}\ \bibnamefont {Nadlinger}}, \bibinfo {author} {\bibfnamefont {P.}~\bibnamefont {Drmota}}, \bibinfo {author} {\bibfnamefont {B.~C.}\ \bibnamefont {Nichol}}, \bibinfo {author} {\bibfnamefont {G.}~\bibnamefont {Araneda}}, \bibinfo {author} {\bibfnamefont {D.}~\bibnamefont {Main}}, \bibinfo {author} {\bibfnamefont {R.}~\bibnamefont {Srinivas}}, \bibinfo {author} {\bibfnamefont {D.~M.}\ \bibnamefont {Lucas}}, \bibinfo {author} {\bibfnamefont {C.~J.}\ \bibnamefont {Ballance}}, \bibinfo {author} {\bibfnamefont {K.}~\bibnamefont {Ivanov}}, \bibinfo {author} {\bibfnamefont {E.~Y.-Z.}\ \bibnamefont {Tan}}, \bibinfo {author} {\bibfnamefont {P.}~\bibnamefont {Sekatski}}, \bibinfo {author} {\bibfnamefont {R.~L.}\ \bibnamefont {Urbanke}}, \bibinfo {author} {\bibfnamefont {R.}~\bibnamefont {Renner}}, \bibinfo {author} {\bibfnamefont {N.}~\bibnamefont {Sangouard}},\ and\ \bibinfo {author} {\bibfnamefont {J.-D.}\ \bibnamefont {Bancal}},\ }\bibfield  {title}
  {\bibinfo {title} {Experimental quantum key distribution certified by bell’s theorem},\ }\href {https://doi.org/10.1038/s41586-022-04941-5} {\bibfield  {journal} {\bibinfo  {journal} {Nature}\ }\textbf {\bibinfo {volume} {607}},\ \bibinfo {pages} {682–686} (\bibinfo {year} {2022})}\BibitemShut {NoStop}%
\bibitem [{\citenamefont {Tan}\ \emph {et~al.}(2022)\citenamefont {Tan}, \citenamefont {Sekatski}, \citenamefont {Bancal}, \citenamefont {Schwonnek}, \citenamefont {Renner}, \citenamefont {Sangouard},\ and\ \citenamefont {Lim}}]{Tan2022}%
  \BibitemOpen
  \bibfield  {author} {\bibinfo {author} {\bibfnamefont {E.~Y.-Z.}\ \bibnamefont {Tan}}, \bibinfo {author} {\bibfnamefont {P.}~\bibnamefont {Sekatski}}, \bibinfo {author} {\bibfnamefont {J.-D.}\ \bibnamefont {Bancal}}, \bibinfo {author} {\bibfnamefont {R.}~\bibnamefont {Schwonnek}}, \bibinfo {author} {\bibfnamefont {R.}~\bibnamefont {Renner}}, \bibinfo {author} {\bibfnamefont {N.}~\bibnamefont {Sangouard}},\ and\ \bibinfo {author} {\bibfnamefont {C.~C.-W.}\ \bibnamefont {Lim}},\ }\bibfield  {title} {\bibinfo {title} {Improved diqkd protocols with finite-size analysis},\ }\href {https://doi.org/10.22331/q-2022-12-22-880} {\bibfield  {journal} {\bibinfo  {journal} {Quantum}\ }\textbf {\bibinfo {volume} {6}},\ \bibinfo {pages} {880} (\bibinfo {year} {2022})}\BibitemShut {NoStop}%
\bibitem [{\citenamefont {Ho}\ \emph {et~al.}(2020)\citenamefont {Ho}, \citenamefont {Sekatski}, \citenamefont {Tan}, \citenamefont {Renner}, \citenamefont {Bancal},\ and\ \citenamefont {Sangouard}}]{Noisy_Preprocessing_Ho20}%
  \BibitemOpen
  \bibfield  {author} {\bibinfo {author} {\bibfnamefont {M.}~\bibnamefont {Ho}}, \bibinfo {author} {\bibfnamefont {P.}~\bibnamefont {Sekatski}}, \bibinfo {author} {\bibfnamefont {E.~Y.-Z.}\ \bibnamefont {Tan}}, \bibinfo {author} {\bibfnamefont {R.}~\bibnamefont {Renner}}, \bibinfo {author} {\bibfnamefont {J.-D.}\ \bibnamefont {Bancal}},\ and\ \bibinfo {author} {\bibfnamefont {N.}~\bibnamefont {Sangouard}},\ }\bibfield  {title} {\bibinfo {title} {Noisy preprocessing facilitates a photonic realization of device-independent quantum key distribution},\ }\href {https://doi.org/10.1103/PhysRevLett.124.230502} {\bibfield  {journal} {\bibinfo  {journal} {Phys. Rev. Lett.}\ }\textbf {\bibinfo {volume} {124}},\ \bibinfo {pages} {230502} (\bibinfo {year} {2020})}\BibitemShut {NoStop}%
\end{thebibliography}%


\begin{thebibliography}{50}%
\makeatletter
\providecommand \@ifxundefined [1]{%
 \@ifx{#1\undefined}
}%
\providecommand \@ifnum [1]{%
 \ifnum #1\expandafter \@firstoftwo
 \else \expandafter \@secondoftwo
 \fi
}%
\providecommand \@ifx [1]{%
 \ifx #1\expandafter \@firstoftwo
 \else \expandafter \@secondoftwo
 \fi
}%
\providecommand \natexlab [1]{#1}%
\providecommand \enquote  [1]{``#1''}%
\providecommand \bibnamefont  [1]{#1}%
\providecommand \bibfnamefont [1]{#1}%
\providecommand \citenamefont [1]{#1}%
\providecommand \href@noop [0]{\@secondoftwo}%
\providecommand \href [0]{\begingroup \@sanitize@url \@href}%
\providecommand \@href[1]{\@@startlink{#1}\@@href}%
\providecommand \@@href[1]{\endgroup#1\@@endlink}%
\providecommand \@sanitize@url [0]{\catcode `\\12\catcode `\$12\catcode `\&12\catcode `\#12\catcode `\^12\catcode `\_12\catcode `\%12\relax}%
\providecommand \@@startlink[1]{}%
\providecommand \@@endlink[0]{}%
\providecommand \url  [0]{\begingroup\@sanitize@url \@url }%
\providecommand \@url [1]{\endgroup\@href {#1}{\urlprefix }}%
\providecommand \urlprefix  [0]{URL }%
\providecommand \Eprint [0]{\href }%
\providecommand \doibase [0]{https://doi.org/}%
\providecommand \selectlanguage [0]{\@gobble}%
\providecommand \bibinfo  [0]{\@secondoftwo}%
\providecommand \bibfield  [0]{\@secondoftwo}%
\providecommand \translation [1]{[#1]}%
\providecommand \BibitemOpen [0]{}%
\providecommand \bibitemStop [0]{}%
\providecommand \bibitemNoStop [0]{.\EOS\space}%
\providecommand \EOS [0]{\spacefactor3000\relax}%
\providecommand \BibitemShut  [1]{\csname bibitem#1\endcsname}%
\let\auto@bib@innerbib\@empty
\bibitem [{\citenamefont {Bennett}\ and\ \citenamefont {Brassard}(1984)}]{Bennett84}%
  \BibitemOpen
  \bibfield  {author} {\bibinfo {author} {\bibfnamefont {C.~H.}\ \bibnamefont {Bennett}}\ and\ \bibinfo {author} {\bibfnamefont {G.}~\bibnamefont {Brassard}},\ }\bibfield  {title} {\bibinfo {title} {Quantum cryptography: public key distribution and coin tossing.},\ }\href@noop {} {\bibfield  {journal} {\bibinfo  {journal} {Proceedings of IEEE International Conference on Computers, Systems, and Signal Processing}\ ,\ \bibinfo {pages} {175}} (\bibinfo {year} {1984})}\BibitemShut {NoStop}%
\bibitem [{\citenamefont {Ekert}(1991)}]{Ekert91}%
  \BibitemOpen
  \bibfield  {author} {\bibinfo {author} {\bibfnamefont {A.~K.}\ \bibnamefont {Ekert}},\ }\bibfield  {title} {\bibinfo {title} {Quantum cryptography based on bell's theorem},\ }\href {https://doi.org/10.1103/physrevlett.67.661} {\bibfield  {journal} {\bibinfo  {journal} {Physical Review Letters}\ }\textbf {\bibinfo {volume} {67}},\ \bibinfo {pages} {661} (\bibinfo {year} {1991})}\BibitemShut {NoStop}%
\bibitem [{\citenamefont {Gisin}\ \emph {et~al.}(2002)\citenamefont {Gisin}, \citenamefont {Ribordy}, \citenamefont {Tittel},\ and\ \citenamefont {Zbinden}}]{Gisin2002}%
  \BibitemOpen
  \bibfield  {author} {\bibinfo {author} {\bibfnamefont {N.}~\bibnamefont {Gisin}}, \bibinfo {author} {\bibfnamefont {G.}~\bibnamefont {Ribordy}}, \bibinfo {author} {\bibfnamefont {W.}~\bibnamefont {Tittel}},\ and\ \bibinfo {author} {\bibfnamefont {H.}~\bibnamefont {Zbinden}},\ }\bibfield  {title} {\bibinfo {title} {Quantum cryptography},\ }\href {https://doi.org/10.1103/RevModPhys.74.145} {\bibfield  {journal} {\bibinfo  {journal} {Reviews of Modern Physics}\ }\textbf {\bibinfo {volume} {74}},\ \bibinfo {pages} {145} (\bibinfo {year} {2002})}\BibitemShut {NoStop}%
\bibitem [{\citenamefont {Lo}\ \emph {et~al.}(2014)\citenamefont {Lo}, \citenamefont {Curty},\ and\ \citenamefont {Tamaki}}]{Lo2014}%
  \BibitemOpen
  \bibfield  {author} {\bibinfo {author} {\bibfnamefont {H.-K.}\ \bibnamefont {Lo}}, \bibinfo {author} {\bibfnamefont {M.}~\bibnamefont {Curty}},\ and\ \bibinfo {author} {\bibfnamefont {K.}~\bibnamefont {Tamaki}},\ }\bibfield  {title} {\bibinfo {title} {Secure quantum key distribution},\ }\href {https://doi.org/10.1038/nphoton.2014.149} {\bibfield  {journal} {\bibinfo  {journal} {Nature Photonics}\ }\textbf {\bibinfo {volume} {8}},\ \bibinfo {pages} {595} (\bibinfo {year} {2014})}\BibitemShut {NoStop}%
\bibitem [{\citenamefont {Scarani}\ \emph {et~al.}(2009)\citenamefont {Scarani}, \citenamefont {Bechmann-Pasquinucci}, \citenamefont {Cerf}, \citenamefont {Dušek}, \citenamefont {Lütkenhaus},\ and\ \citenamefont {Peev}}]{Scarani2009}%
  \BibitemOpen
  \bibfield  {author} {\bibinfo {author} {\bibfnamefont {V.}~\bibnamefont {Scarani}}, \bibinfo {author} {\bibfnamefont {H.}~\bibnamefont {Bechmann-Pasquinucci}}, \bibinfo {author} {\bibfnamefont {N.~J.}\ \bibnamefont {Cerf}}, \bibinfo {author} {\bibfnamefont {M.}~\bibnamefont {Dušek}}, \bibinfo {author} {\bibfnamefont {N.}~\bibnamefont {Lütkenhaus}},\ and\ \bibinfo {author} {\bibfnamefont {M.}~\bibnamefont {Peev}},\ }\bibfield  {title} {\bibinfo {title} {The security of practical quantum key distribution},\ }\href {https://doi.org/10.1103/RevModPhys.81.1301} {\bibfield  {journal} {\bibinfo  {journal} {Reviews of Modern Physics}\ }\textbf {\bibinfo {volume} {81}},\ \bibinfo {pages} {1301} (\bibinfo {year} {2009})}\BibitemShut {NoStop}%
\bibitem [{\citenamefont {Gerhardt}\ \emph {et~al.}(2011)\citenamefont {Gerhardt}, \citenamefont {Liu}, \citenamefont {Lamas-Linares}, \citenamefont {Skaar}, \citenamefont {Kurtsiefer},\ and\ \citenamefont {Makarov}}]{Gerhardt2011}%
  \BibitemOpen
  \bibfield  {author} {\bibinfo {author} {\bibfnamefont {I.}~\bibnamefont {Gerhardt}}, \bibinfo {author} {\bibfnamefont {Q.}~\bibnamefont {Liu}}, \bibinfo {author} {\bibfnamefont {A.}~\bibnamefont {Lamas-Linares}}, \bibinfo {author} {\bibfnamefont {J.}~\bibnamefont {Skaar}}, \bibinfo {author} {\bibfnamefont {C.}~\bibnamefont {Kurtsiefer}},\ and\ \bibinfo {author} {\bibfnamefont {V.}~\bibnamefont {Makarov}},\ }\bibfield  {title} {\bibinfo {title} {Full-field implementation of a perfect eavesdropper on a quantum cryptography system},\ }\href {https://doi.org/10.1038/ncomms1348} {\bibfield  {journal} {\bibinfo  {journal} {Nature Communications}\ }\textbf {\bibinfo {volume} {2}},\ \bibinfo {pages} {1348} (\bibinfo {year} {2011})}\BibitemShut {NoStop}%
\bibitem [{\citenamefont {Lydersen}\ \emph {et~al.}(2010)\citenamefont {Lydersen}, \citenamefont {Wiechers}, \citenamefont {Wittmann}, \citenamefont {Elser}, \citenamefont {Skaar},\ and\ \citenamefont {Makarov}}]{Lydersen2010}%
  \BibitemOpen
  \bibfield  {author} {\bibinfo {author} {\bibfnamefont {L.}~\bibnamefont {Lydersen}}, \bibinfo {author} {\bibfnamefont {C.}~\bibnamefont {Wiechers}}, \bibinfo {author} {\bibfnamefont {C.}~\bibnamefont {Wittmann}}, \bibinfo {author} {\bibfnamefont {D.}~\bibnamefont {Elser}}, \bibinfo {author} {\bibfnamefont {J.}~\bibnamefont {Skaar}},\ and\ \bibinfo {author} {\bibfnamefont {V.}~\bibnamefont {Makarov}},\ }\bibfield  {title} {\bibinfo {title} {Hacking commercial quantum cryptography systems by tailored bright illumination},\ }\href {https://doi.org/10.1038/nphoton.2010.214} {\bibfield  {journal} {\bibinfo  {journal} {Nature Photonics}\ }\textbf {\bibinfo {volume} {4}},\ \bibinfo {pages} {686} (\bibinfo {year} {2010})}\BibitemShut {NoStop}%
\bibitem [{\citenamefont {Zhao}\ \emph {et~al.}(2008)\citenamefont {Zhao}, \citenamefont {Fung}, \citenamefont {Qi}, \citenamefont {Chen},\ and\ \citenamefont {Lo}}]{Zhao2008}%
  \BibitemOpen
  \bibfield  {author} {\bibinfo {author} {\bibfnamefont {Y.}~\bibnamefont {Zhao}}, \bibinfo {author} {\bibfnamefont {C.-H.~F.}\ \bibnamefont {Fung}}, \bibinfo {author} {\bibfnamefont {B.}~\bibnamefont {Qi}}, \bibinfo {author} {\bibfnamefont {C.}~\bibnamefont {Chen}},\ and\ \bibinfo {author} {\bibfnamefont {H.-K.}\ \bibnamefont {Lo}},\ }\bibfield  {title} {\bibinfo {title} {Quantum hacking: Experimental demonstration of time-shift attack against practical quantum-key-distribution systems},\ }\href {https://doi.org/10.1103/PhysRevA.78.042333} {\bibfield  {journal} {\bibinfo  {journal} {Physical Review A}\ }\textbf {\bibinfo {volume} {78}},\ \bibinfo {pages} {042333} (\bibinfo {year} {2008})}\BibitemShut {NoStop}%
\bibitem [{\citenamefont {Weier}\ \emph {et~al.}(2011)\citenamefont {Weier}, \citenamefont {Krauss}, \citenamefont {Rau}, \citenamefont {Fürst}, \citenamefont {Nauerth},\ and\ \citenamefont {Weinfurter}}]{Weier2011}%
  \BibitemOpen
  \bibfield  {author} {\bibinfo {author} {\bibfnamefont {H.}~\bibnamefont {Weier}}, \bibinfo {author} {\bibfnamefont {H.}~\bibnamefont {Krauss}}, \bibinfo {author} {\bibfnamefont {M.}~\bibnamefont {Rau}}, \bibinfo {author} {\bibfnamefont {M.}~\bibnamefont {Fürst}}, \bibinfo {author} {\bibfnamefont {S.}~\bibnamefont {Nauerth}},\ and\ \bibinfo {author} {\bibfnamefont {H.}~\bibnamefont {Weinfurter}},\ }\bibfield  {title} {\bibinfo {title} {Quantum eavesdropping without interception: An attack exploiting the dead time of single-photon detectors},\ }\href {https://doi.org/10.1088/1367-2630/13/7/073024} {\bibfield  {journal} {\bibinfo  {journal} {New Journal of Physics}\ }\textbf {\bibinfo {volume} {13}},\ \bibinfo {pages} {073024} (\bibinfo {year} {2011})}\BibitemShut {NoStop}%
\bibitem [{\citenamefont {Mayers}\ and\ \citenamefont {Yao}(2004)}]{Mayers2004}%
  \BibitemOpen
  \bibfield  {author} {\bibinfo {author} {\bibfnamefont {D.}~\bibnamefont {Mayers}}\ and\ \bibinfo {author} {\bibfnamefont {A.}~\bibnamefont {Yao}},\ }\bibfield  {title} {\bibinfo {title} {Self testing quantum apparatus},\ }\href@noop {} {\bibfield  {journal} {\bibinfo  {journal} {Quantum Information and Computation}\ }\textbf {\bibinfo {volume} {4}},\ \bibinfo {pages} {273} (\bibinfo {year} {2004})}\BibitemShut {NoStop}%
\bibitem [{\citenamefont {Acín}\ \emph {et~al.}(2007)\citenamefont {Acín}, \citenamefont {Brunner}, \citenamefont {Gisin}, \citenamefont {Massar}, \citenamefont {Pironio},\ and\ \citenamefont {Scarani}}]{Acin2007}%
  \BibitemOpen
  \bibfield  {author} {\bibinfo {author} {\bibfnamefont {A.}~\bibnamefont {Acín}}, \bibinfo {author} {\bibfnamefont {N.}~\bibnamefont {Brunner}}, \bibinfo {author} {\bibfnamefont {N.}~\bibnamefont {Gisin}}, \bibinfo {author} {\bibfnamefont {S.}~\bibnamefont {Massar}}, \bibinfo {author} {\bibfnamefont {S.}~\bibnamefont {Pironio}},\ and\ \bibinfo {author} {\bibfnamefont {V.}~\bibnamefont {Scarani}},\ }\bibfield  {title} {\bibinfo {title} {Device-independent security of quantum cryptography against collective attacks},\ }\href {https://doi.org/10.1103/PhysRevLett.98.230501} {\bibfield  {journal} {\bibinfo  {journal} {Physical Review Letters}\ }\textbf {\bibinfo {volume} {98}},\ \bibinfo {pages} {230501} (\bibinfo {year} {2007})}\BibitemShut {NoStop}%
\bibitem [{\citenamefont {Vazirani}\ and\ \citenamefont {Vidick}(2014)}]{Vazirani2014}%
  \BibitemOpen
  \bibfield  {author} {\bibinfo {author} {\bibfnamefont {U.}~\bibnamefont {Vazirani}}\ and\ \bibinfo {author} {\bibfnamefont {T.}~\bibnamefont {Vidick}},\ }\bibfield  {title} {\bibinfo {title} {Fully device-independent quantum key distribution},\ }\href {https://doi.org/10.1103/PhysRevLett.113.140501} {\bibfield  {journal} {\bibinfo  {journal} {Physical Review Letters}\ }\textbf {\bibinfo {volume} {113}},\ \bibinfo {pages} {140501} (\bibinfo {year} {2014})}\BibitemShut {NoStop}%
\bibitem [{\citenamefont {Ac\'{\i}n}\ \emph {et~al.}(2016)\citenamefont {Ac\'{\i}n}, \citenamefont {Cavalcanti}, \citenamefont {Passaro}, \citenamefont {Pironio},\ and\ \citenamefont {Skrzypczyk}}]{Acin2016}%
  \BibitemOpen
  \bibfield  {author} {\bibinfo {author} {\bibfnamefont {A.}~\bibnamefont {Ac\'{\i}n}}, \bibinfo {author} {\bibfnamefont {D.}~\bibnamefont {Cavalcanti}}, \bibinfo {author} {\bibfnamefont {E.}~\bibnamefont {Passaro}}, \bibinfo {author} {\bibfnamefont {S.}~\bibnamefont {Pironio}},\ and\ \bibinfo {author} {\bibfnamefont {P.}~\bibnamefont {Skrzypczyk}},\ }\bibfield  {title} {\bibinfo {title} {Necessary detection efficiencies for secure quantum key distribution and bound randomness},\ }\href {https://doi.org/10.1103/PhysRevA.93.012319} {\bibfield  {journal} {\bibinfo  {journal} {Phys. Rev. A}\ }\textbf {\bibinfo {volume} {93}},\ \bibinfo {pages} {012319} (\bibinfo {year} {2016})}\BibitemShut {NoStop}%
\bibitem [{\citenamefont {Nadlinger}\ \emph {et~al.}(2022)\citenamefont {Nadlinger}, \citenamefont {Drmota}, \citenamefont {Nichol}, \citenamefont {Araneda}, \citenamefont {Main}, \citenamefont {Srinivas}, \citenamefont {Lucas}, \citenamefont {Ballance}, \citenamefont {Ivanov}, \citenamefont {Tan}, \citenamefont {Sekatski}, \citenamefont {Urbanke}, \citenamefont {Renner}, \citenamefont {Sangouard},\ and\ \citenamefont {Bancal}}]{Nadlinger_2022}%
  \BibitemOpen
  \bibfield  {author} {\bibinfo {author} {\bibfnamefont {D.~P.}\ \bibnamefont {Nadlinger}}, \bibinfo {author} {\bibfnamefont {P.}~\bibnamefont {Drmota}}, \bibinfo {author} {\bibfnamefont {B.~C.}\ \bibnamefont {Nichol}}, \bibinfo {author} {\bibfnamefont {G.}~\bibnamefont {Araneda}}, \bibinfo {author} {\bibfnamefont {D.}~\bibnamefont {Main}}, \bibinfo {author} {\bibfnamefont {R.}~\bibnamefont {Srinivas}}, \bibinfo {author} {\bibfnamefont {D.~M.}\ \bibnamefont {Lucas}}, \bibinfo {author} {\bibfnamefont {C.~J.}\ \bibnamefont {Ballance}}, \bibinfo {author} {\bibfnamefont {K.}~\bibnamefont {Ivanov}}, \bibinfo {author} {\bibfnamefont {E.~Y.-Z.}\ \bibnamefont {Tan}}, \bibinfo {author} {\bibfnamefont {P.}~\bibnamefont {Sekatski}}, \bibinfo {author} {\bibfnamefont {R.~L.}\ \bibnamefont {Urbanke}}, \bibinfo {author} {\bibfnamefont {R.}~\bibnamefont {Renner}}, \bibinfo {author} {\bibfnamefont {N.}~\bibnamefont {Sangouard}},\ and\ \bibinfo {author} {\bibfnamefont {J.-D.}\ \bibnamefont {Bancal}},\ }\bibfield  {title}
  {\bibinfo {title} {Experimental quantum key distribution certified by bell’s theorem},\ }\href {https://doi.org/10.1038/s41586-022-04941-5} {\bibfield  {journal} {\bibinfo  {journal} {Nature}\ }\textbf {\bibinfo {volume} {607}},\ \bibinfo {pages} {682–686} (\bibinfo {year} {2022})}\BibitemShut {NoStop}%
\bibitem [{\citenamefont {Zhang}\ \emph {et~al.}(2022)\citenamefont {Zhang}, \citenamefont {van Leent}, \citenamefont {Redeker}, \citenamefont {Garthoff}, \citenamefont {Schwonnek}, \citenamefont {Fertig}, \citenamefont {Eppelt}, \citenamefont {Rosenfeld}, \citenamefont {Scarani}, \citenamefont {Lim},\ and\ \citenamefont {Weinfurter}}]{Zhang2022}%
  \BibitemOpen
  \bibfield  {author} {\bibinfo {author} {\bibfnamefont {W.}~\bibnamefont {Zhang}}, \bibinfo {author} {\bibfnamefont {T.}~\bibnamefont {van Leent}}, \bibinfo {author} {\bibfnamefont {K.}~\bibnamefont {Redeker}}, \bibinfo {author} {\bibfnamefont {R.}~\bibnamefont {Garthoff}}, \bibinfo {author} {\bibfnamefont {R.}~\bibnamefont {Schwonnek}}, \bibinfo {author} {\bibfnamefont {F.}~\bibnamefont {Fertig}}, \bibinfo {author} {\bibfnamefont {S.}~\bibnamefont {Eppelt}}, \bibinfo {author} {\bibfnamefont {W.}~\bibnamefont {Rosenfeld}}, \bibinfo {author} {\bibfnamefont {V.}~\bibnamefont {Scarani}}, \bibinfo {author} {\bibfnamefont {C.~C.-W.}\ \bibnamefont {Lim}},\ and\ \bibinfo {author} {\bibfnamefont {H.}~\bibnamefont {Weinfurter}},\ }\bibfield  {title} {\bibinfo {title} {A device-independent quantum key distribution system for distant users},\ }\href {https://doi.org/10.1038/s41586-022-04891-y} {\bibfield  {journal} {\bibinfo  {journal} {Nature}\ }\textbf {\bibinfo {volume} {607}},\ \bibinfo {pages} {687–691}
  (\bibinfo {year} {2022})}\BibitemShut {NoStop}%
\bibitem [{\citenamefont {Zapatero}\ \emph {et~al.}(2023)\citenamefont {Zapatero}, \citenamefont {van Leent}, \citenamefont {Arnon-Friedman}, \citenamefont {Liu}, \citenamefont {Zhang}, \citenamefont {Weinfurter},\ and\ \citenamefont {Curty}}]{Zapatero2023}%
  \BibitemOpen
  \bibfield  {author} {\bibinfo {author} {\bibfnamefont {V.}~\bibnamefont {Zapatero}}, \bibinfo {author} {\bibfnamefont {T.}~\bibnamefont {van Leent}}, \bibinfo {author} {\bibfnamefont {R.}~\bibnamefont {Arnon-Friedman}}, \bibinfo {author} {\bibfnamefont {W.-Z.}\ \bibnamefont {Liu}}, \bibinfo {author} {\bibfnamefont {Q.}~\bibnamefont {Zhang}}, \bibinfo {author} {\bibfnamefont {H.}~\bibnamefont {Weinfurter}},\ and\ \bibinfo {author} {\bibfnamefont {M.}~\bibnamefont {Curty}},\ }\bibfield  {title} {\bibinfo {title} {Advances in device-independent quantum key distribution},\ }\bibfield  {journal} {\bibinfo  {journal} {npj Quantum Information}\ }\textbf {\bibinfo {volume} {9}},\ \href {https://doi.org/10.1038/s41534-023-00684-x} {10.1038/s41534-023-00684-x} (\bibinfo {year} {2023})\BibitemShut {NoStop}%
\bibitem [{\citenamefont {Valcarce}(2023)}]{Valcarce2023}%
  \BibitemOpen
  \bibfield  {author} {\bibinfo {author} {\bibfnamefont {X.}~\bibnamefont {Valcarce}},\ }\emph {\bibinfo {title} {{Device-independent certification: quantum resources and quantum key distribution}}},\ \href {https://theses.hal.science/tel-04132704} {\bibinfo {type} {Theses}},\ \bibinfo  {school} {{Universit{\'e} Paris-Saclay}} (\bibinfo {year} {2023})\BibitemShut {NoStop}%
\bibitem [{\citenamefont {Gisin}\ \emph {et~al.}(2010)\citenamefont {Gisin}, \citenamefont {Pironio},\ and\ \citenamefont {Sangouard}}]{Gisin10}%
  \BibitemOpen
  \bibfield  {author} {\bibinfo {author} {\bibfnamefont {N.}~\bibnamefont {Gisin}}, \bibinfo {author} {\bibfnamefont {S.}~\bibnamefont {Pironio}},\ and\ \bibinfo {author} {\bibfnamefont {N.}~\bibnamefont {Sangouard}},\ }\bibfield  {title} {\bibinfo {title} {Proposal for implementing device-independent quantum key distribution based on a heralded qubit amplifier},\ }\href {https://doi.org/10.1103/PhysRevLett.105.070501} {\bibfield  {journal} {\bibinfo  {journal} {Phys. Rev. Lett.}\ }\textbf {\bibinfo {volume} {105}},\ \bibinfo {pages} {070501} (\bibinfo {year} {2010})}\BibitemShut {NoStop}%
\bibitem [{\citenamefont {Ko{\l{}}ody{\'{n}}ski}\ \emph {et~al.}(2020)\citenamefont {Ko{\l{}}ody{\'{n}}ski}, \citenamefont {M{\'{a}}ttar}, \citenamefont {Skrzypczyk}, \citenamefont {Woodhead}, \citenamefont {Cavalcanti}, \citenamefont {Banaszek},\ and\ \citenamefont {Ac{\'{i}}n}}]{Kolodynski2020}%
  \BibitemOpen
  \bibfield  {author} {\bibinfo {author} {\bibfnamefont {J.}~\bibnamefont {Ko{\l{}}ody{\'{n}}ski}}, \bibinfo {author} {\bibfnamefont {A.}~\bibnamefont {M{\'{a}}ttar}}, \bibinfo {author} {\bibfnamefont {P.}~\bibnamefont {Skrzypczyk}}, \bibinfo {author} {\bibfnamefont {E.}~\bibnamefont {Woodhead}}, \bibinfo {author} {\bibfnamefont {D.}~\bibnamefont {Cavalcanti}}, \bibinfo {author} {\bibfnamefont {K.}~\bibnamefont {Banaszek}},\ and\ \bibinfo {author} {\bibfnamefont {A.}~\bibnamefont {Ac{\'{i}}n}},\ }\bibfield  {title} {\bibinfo {title} {Device-independent quantum key distribution with single-photon sources},\ }\href {https://doi.org/10.22331/q-2020-04-30-260} {\bibfield  {journal} {\bibinfo  {journal} {{Quantum}}\ }\textbf {\bibinfo {volume} {4}},\ \bibinfo {pages} {260} (\bibinfo {year} {2020})}\BibitemShut {NoStop}%
\bibitem [{\citenamefont {Gonz\'{a}lez-Ruiz}\ \emph {et~al.}(2024)\citenamefont {Gonz\'{a}lez-Ruiz}, \citenamefont {Rivera-Dean}, \citenamefont {Cenni}, \citenamefont {S{\o}rensen}, \citenamefont {Ac\'{i}n},\ and\ \citenamefont {Oudot}}]{Gonzalez-Ruiz24}%
  \BibitemOpen
  \bibfield  {author} {\bibinfo {author} {\bibfnamefont {E.~M.}\ \bibnamefont {Gonz\'{a}lez-Ruiz}}, \bibinfo {author} {\bibfnamefont {J.}~\bibnamefont {Rivera-Dean}}, \bibinfo {author} {\bibfnamefont {M.~F.~B.}\ \bibnamefont {Cenni}}, \bibinfo {author} {\bibfnamefont {A.~S.}\ \bibnamefont {S{\o}rensen}}, \bibinfo {author} {\bibfnamefont {A.}~\bibnamefont {Ac\'{i}n}},\ and\ \bibinfo {author} {\bibfnamefont {E.}~\bibnamefont {Oudot}},\ }\bibfield  {title} {\bibinfo {title} {Device-independent quantum key distribution with realistic single-photon source implementations},\ }\href {https://doi.org/10.1364/OE.497935} {\bibfield  {journal} {\bibinfo  {journal} {Opt. Express}\ }\textbf {\bibinfo {volume} {32}},\ \bibinfo {pages} {13181} (\bibinfo {year} {2024})}\BibitemShut {NoStop}%
\bibitem [{\citenamefont {Mycroft}\ \emph {et~al.}(2023)\citenamefont {Mycroft}, \citenamefont {McDermott}, \citenamefont {Buraczewski},\ and\ \citenamefont {Stobi\ifmmode~\acute{n}\else \'{n}\fi{}ska}}]{Mycroft23}%
  \BibitemOpen
  \bibfield  {author} {\bibinfo {author} {\bibfnamefont {M.~E.}\ \bibnamefont {Mycroft}}, \bibinfo {author} {\bibfnamefont {T.}~\bibnamefont {McDermott}}, \bibinfo {author} {\bibfnamefont {A.}~\bibnamefont {Buraczewski}},\ and\ \bibinfo {author} {\bibfnamefont {M.}~\bibnamefont {Stobi\ifmmode~\acute{n}\else \'{n}\fi{}ska}},\ }\bibfield  {title} {\bibinfo {title} {Proposal for the distribution of multiphoton entanglement with optimal rate-distance scaling},\ }\href {https://doi.org/10.1103/PhysRevA.107.012607} {\bibfield  {journal} {\bibinfo  {journal} {Phys. Rev. A}\ }\textbf {\bibinfo {volume} {107}},\ \bibinfo {pages} {012607} (\bibinfo {year} {2023})}\BibitemShut {NoStop}%
\bibitem [{\citenamefont {Xie}\ \emph {et~al.}(2021)\citenamefont {Xie}, \citenamefont {Li}, \citenamefont {Lu}, \citenamefont {Cao}, \citenamefont {Liu}, \citenamefont {Yin},\ and\ \citenamefont {Chen}}]{Xie2021}%
  \BibitemOpen
  \bibfield  {author} {\bibinfo {author} {\bibfnamefont {Y.-M.}\ \bibnamefont {Xie}}, \bibinfo {author} {\bibfnamefont {B.-H.}\ \bibnamefont {Li}}, \bibinfo {author} {\bibfnamefont {Y.-S.}\ \bibnamefont {Lu}}, \bibinfo {author} {\bibfnamefont {X.-Y.}\ \bibnamefont {Cao}}, \bibinfo {author} {\bibfnamefont {W.-B.}\ \bibnamefont {Liu}}, \bibinfo {author} {\bibfnamefont {H.-L.}\ \bibnamefont {Yin}},\ and\ \bibinfo {author} {\bibfnamefont {Z.-B.}\ \bibnamefont {Chen}},\ }\bibfield  {title} {\bibinfo {title} {Overcoming the rate--distance limit of device-independent quantum key distribution},\ }\href {https://doi.org/10.1364/OL.417851} {\bibfield  {journal} {\bibinfo  {journal} {Opt. Lett.}\ }\textbf {\bibinfo {volume} {46}},\ \bibinfo {pages} {1632} (\bibinfo {year} {2021})}\BibitemShut {NoStop}%
\bibitem [{\citenamefont {Lucamarini}\ \emph {et~al.}(2018)\citenamefont {Lucamarini}, \citenamefont {Yuan}, \citenamefont {Dynes},\ and\ \citenamefont {Shields}}]{Lucamarini2018}%
  \BibitemOpen
  \bibfield  {author} {\bibinfo {author} {\bibfnamefont {M.}~\bibnamefont {Lucamarini}}, \bibinfo {author} {\bibfnamefont {Z.~L.}\ \bibnamefont {Yuan}}, \bibinfo {author} {\bibfnamefont {J.~F.}\ \bibnamefont {Dynes}},\ and\ \bibinfo {author} {\bibfnamefont {A.~J.}\ \bibnamefont {Shields}},\ }\bibfield  {title} {\bibinfo {title} {Overcoming the rate--distance limit of quantum key distribution without quantum repeaters},\ }\href {https://doi.org/10.1038/s41586-018-0066-6} {\bibfield  {journal} {\bibinfo  {journal} {Nature}\ }\textbf {\bibinfo {volume} {557}},\ \bibinfo {pages} {400} (\bibinfo {year} {2018})}\BibitemShut {NoStop}%
\bibitem [{\citenamefont {Tan}\ \emph {et~al.}(2022)\citenamefont {Tan}, \citenamefont {Sekatski}, \citenamefont {Bancal}, \citenamefont {Schwonnek}, \citenamefont {Renner}, \citenamefont {Sangouard},\ and\ \citenamefont {Lim}}]{Tan2022}%
  \BibitemOpen
  \bibfield  {author} {\bibinfo {author} {\bibfnamefont {E.~Y.-Z.}\ \bibnamefont {Tan}}, \bibinfo {author} {\bibfnamefont {P.}~\bibnamefont {Sekatski}}, \bibinfo {author} {\bibfnamefont {J.-D.}\ \bibnamefont {Bancal}}, \bibinfo {author} {\bibfnamefont {R.}~\bibnamefont {Schwonnek}}, \bibinfo {author} {\bibfnamefont {R.}~\bibnamefont {Renner}}, \bibinfo {author} {\bibfnamefont {N.}~\bibnamefont {Sangouard}},\ and\ \bibinfo {author} {\bibfnamefont {C.~C.-W.}\ \bibnamefont {Lim}},\ }\bibfield  {title} {\bibinfo {title} {Improved diqkd protocols with finite-size analysis},\ }\href {https://doi.org/10.22331/q-2022-12-22-880} {\bibfield  {journal} {\bibinfo  {journal} {Quantum}\ }\textbf {\bibinfo {volume} {6}},\ \bibinfo {pages} {880} (\bibinfo {year} {2022})}\BibitemShut {NoStop}%
\bibitem [{Note1()}]{Note1}%
  \BibitemOpen
  \bibinfo {note} {Note that a detection on the right detector will herald the state $(\ket {10}-\ket {01})/\protect \sqrt {2}$}\BibitemShut {NoStop}%
\bibitem [{SM()}]{SM}%
  \BibitemOpen
  \href@noop {} {}\bibinfo {note} {See Supplementary Material at [URL] for a deeper description of the measurements, the heralded state with ideal sources, as well as an analysis with realistic sources, a dark count description, and the finite-size analysis, which includes Ref. \cite{CapraraVivoli2015,SPDC,Christ2011,Muljarov2004,Zhao2020,Gonzalez2022}.}\BibitemShut {Stop}%
\bibitem [{\citenamefont {Banaszek}\ and\ \citenamefont {Wódkiewicz}(1999)}]{Banaszek_1999}%
  \BibitemOpen
  \bibfield  {author} {\bibinfo {author} {\bibfnamefont {K.}~\bibnamefont {Banaszek}}\ and\ \bibinfo {author} {\bibfnamefont {K.}~\bibnamefont {Wódkiewicz}},\ }\bibfield  {title} {\bibinfo {title} {Testing quantum nonlocality in phase space},\ }\href {https://doi.org/10.1103/physrevlett.82.2009} {\bibfield  {journal} {\bibinfo  {journal} {Physical Review Letters}\ }\textbf {\bibinfo {volume} {82}},\ \bibinfo {pages} {2009–2013} (\bibinfo {year} {1999})}\BibitemShut {NoStop}%
\bibitem [{\citenamefont {Caspar}\ \emph {et~al.}(2020)\citenamefont {Caspar}, \citenamefont {Verbanis}, \citenamefont {Oudot}, \citenamefont {Maring}, \citenamefont {Samara}, \citenamefont {Caloz}, \citenamefont {Perrenoud}, \citenamefont {Sekatski}, \citenamefont {Martin}, \citenamefont {Sangouard}, \citenamefont {Zbinden},\ and\ \citenamefont {Thew}}]{Caspar2020}%
  \BibitemOpen
  \bibfield  {author} {\bibinfo {author} {\bibfnamefont {P.}~\bibnamefont {Caspar}}, \bibinfo {author} {\bibfnamefont {E.}~\bibnamefont {Verbanis}}, \bibinfo {author} {\bibfnamefont {E.}~\bibnamefont {Oudot}}, \bibinfo {author} {\bibfnamefont {N.}~\bibnamefont {Maring}}, \bibinfo {author} {\bibfnamefont {F.}~\bibnamefont {Samara}}, \bibinfo {author} {\bibfnamefont {M.}~\bibnamefont {Caloz}}, \bibinfo {author} {\bibfnamefont {M.}~\bibnamefont {Perrenoud}}, \bibinfo {author} {\bibfnamefont {P.}~\bibnamefont {Sekatski}}, \bibinfo {author} {\bibfnamefont {A.}~\bibnamefont {Martin}}, \bibinfo {author} {\bibfnamefont {N.}~\bibnamefont {Sangouard}}, \bibinfo {author} {\bibfnamefont {H.}~\bibnamefont {Zbinden}},\ and\ \bibinfo {author} {\bibfnamefont {R.~T.}\ \bibnamefont {Thew}},\ }\bibfield  {title} {\bibinfo {title} {Heralded distribution of single-photon path entanglement},\ }\href {https://doi.org/10.1103/PhysRevLett.125.110506} {\bibfield  {journal} {\bibinfo  {journal} {Physical Review Letters}\ }\textbf
  {\bibinfo {volume} {125}},\ \bibinfo {pages} {110506} (\bibinfo {year} {2020})}\BibitemShut {NoStop}%
\bibitem [{\citenamefont {Paris}(1996)}]{Paris1996DisplacementOB}%
  \BibitemOpen
  \bibfield  {author} {\bibinfo {author} {\bibfnamefont {M.~G.}\ \bibnamefont {Paris}},\ }\bibfield  {title} {\bibinfo {title} {Displacement operator by beam splitter},\ }\href {https://api.semanticscholar.org/CorpusID:120054739} {\bibfield  {journal} {\bibinfo  {journal} {Physics Letters A}\ }\textbf {\bibinfo {volume} {217}},\ \bibinfo {pages} {78} (\bibinfo {year} {1996})}\BibitemShut {NoStop}%
\bibitem [{\citenamefont {Clauser}\ \emph {et~al.}(1969)\citenamefont {Clauser}, \citenamefont {Horne}, \citenamefont {Shimony},\ and\ \citenamefont {Holt}}]{CHSH}%
  \BibitemOpen
  \bibfield  {author} {\bibinfo {author} {\bibfnamefont {J.~F.}\ \bibnamefont {Clauser}}, \bibinfo {author} {\bibfnamefont {M.~A.}\ \bibnamefont {Horne}}, \bibinfo {author} {\bibfnamefont {A.}~\bibnamefont {Shimony}},\ and\ \bibinfo {author} {\bibfnamefont {R.~A.}\ \bibnamefont {Holt}},\ }\bibfield  {title} {\bibinfo {title} {Proposed experiment to test local hidden-variable theories},\ }\href {https://doi.org/10.1103/PhysRevLett.23.880} {\bibfield  {journal} {\bibinfo  {journal} {Phys. Rev. Lett.}\ }\textbf {\bibinfo {volume} {23}},\ \bibinfo {pages} {880} (\bibinfo {year} {1969})}\BibitemShut {NoStop}%
\bibitem [{\citenamefont {Devetak}\ and\ \citenamefont {Winter}(2005)}]{devetak2005distillation}%
  \BibitemOpen
  \bibfield  {author} {\bibinfo {author} {\bibfnamefont {I.}~\bibnamefont {Devetak}}\ and\ \bibinfo {author} {\bibfnamefont {A.}~\bibnamefont {Winter}},\ }\bibfield  {title} {\bibinfo {title} {Distillation of secret key and entanglement from quantum states},\ }\href@noop {} {\bibfield  {journal} {\bibinfo  {journal} {Proceedings of the Royal Society A: Mathematical, Physical and engineering sciences}\ }\textbf {\bibinfo {volume} {461}},\ \bibinfo {pages} {207} (\bibinfo {year} {2005})}\BibitemShut {NoStop}%
\bibitem [{\citenamefont {Ho}\ \emph {et~al.}(2020)\citenamefont {Ho}, \citenamefont {Sekatski}, \citenamefont {Tan}, \citenamefont {Renner}, \citenamefont {Bancal},\ and\ \citenamefont {Sangouard}}]{Noisy_Preprocessing_Ho20}%
  \BibitemOpen
  \bibfield  {author} {\bibinfo {author} {\bibfnamefont {M.}~\bibnamefont {Ho}}, \bibinfo {author} {\bibfnamefont {P.}~\bibnamefont {Sekatski}}, \bibinfo {author} {\bibfnamefont {E.~Y.-Z.}\ \bibnamefont {Tan}}, \bibinfo {author} {\bibfnamefont {R.}~\bibnamefont {Renner}}, \bibinfo {author} {\bibfnamefont {J.-D.}\ \bibnamefont {Bancal}},\ and\ \bibinfo {author} {\bibfnamefont {N.}~\bibnamefont {Sangouard}},\ }\bibfield  {title} {\bibinfo {title} {Noisy preprocessing facilitates a photonic realization of device-independent quantum key distribution},\ }\href {https://doi.org/10.1103/PhysRevLett.124.230502} {\bibfield  {journal} {\bibinfo  {journal} {Phys. Rev. Lett.}\ }\textbf {\bibinfo {volume} {124}},\ \bibinfo {pages} {230502} (\bibinfo {year} {2020})}\BibitemShut {NoStop}%
\bibitem [{\citenamefont {Dupuis}\ \emph {et~al.}(2020)\citenamefont {Dupuis}, \citenamefont {Fawzi},\ and\ \citenamefont {Renner}}]{Dupuis_2020}%
  \BibitemOpen
  \bibfield  {author} {\bibinfo {author} {\bibfnamefont {F.}~\bibnamefont {Dupuis}}, \bibinfo {author} {\bibfnamefont {O.}~\bibnamefont {Fawzi}},\ and\ \bibinfo {author} {\bibfnamefont {R.}~\bibnamefont {Renner}},\ }\bibfield  {title} {\bibinfo {title} {Entropy accumulation},\ }\href {https://doi.org/10.1007/s00220-020-03839-5} {\bibfield  {journal} {\bibinfo  {journal} {Communications in Mathematical Physics}\ }\textbf {\bibinfo {volume} {379}},\ \bibinfo {pages} {867–913} (\bibinfo {year} {2020})}\BibitemShut {NoStop}%
\bibitem [{\citenamefont {Tomm}\ \emph {et~al.}(2021)\citenamefont {Tomm}, \citenamefont {Javadi}, \citenamefont {Antoniadis}, \citenamefont {Najer}, \citenamefont {Löbl}, \citenamefont {Korsch}, \citenamefont {Schott}, \citenamefont {Valentin}, \citenamefont {Wieck}, \citenamefont {Ludwig},\ and\ \citenamefont {Warburton}}]{Tomm_2021}%
  \BibitemOpen
  \bibfield  {author} {\bibinfo {author} {\bibfnamefont {N.}~\bibnamefont {Tomm}}, \bibinfo {author} {\bibfnamefont {A.}~\bibnamefont {Javadi}}, \bibinfo {author} {\bibfnamefont {N.~O.}\ \bibnamefont {Antoniadis}}, \bibinfo {author} {\bibfnamefont {D.}~\bibnamefont {Najer}}, \bibinfo {author} {\bibfnamefont {M.~C.}\ \bibnamefont {Löbl}}, \bibinfo {author} {\bibfnamefont {A.~R.}\ \bibnamefont {Korsch}}, \bibinfo {author} {\bibfnamefont {R.}~\bibnamefont {Schott}}, \bibinfo {author} {\bibfnamefont {S.~R.}\ \bibnamefont {Valentin}}, \bibinfo {author} {\bibfnamefont {A.~D.}\ \bibnamefont {Wieck}}, \bibinfo {author} {\bibfnamefont {A.}~\bibnamefont {Ludwig}},\ and\ \bibinfo {author} {\bibfnamefont {R.~J.}\ \bibnamefont {Warburton}},\ }\bibfield  {title} {\bibinfo {title} {A bright and fast source of coherent single photons},\ }\href {https://doi.org/10.1038/s41565-020-00831-x} {\bibfield  {journal} {\bibinfo  {journal} {Nature Nanotechnology}\ }\textbf {\bibinfo {volume} {16}},\ \bibinfo {pages} {399–403}
  (\bibinfo {year} {2021})}\BibitemShut {NoStop}%
\bibitem [{\citenamefont {Uppu}\ \emph {et~al.}(2020)\citenamefont {Uppu}, \citenamefont {Pedersen}, \citenamefont {Wang}, \citenamefont {Olesen}, \citenamefont {Papon}, \citenamefont {Zhou}, \citenamefont {Midolo}, \citenamefont {Scholz}, \citenamefont {Wieck}, \citenamefont {Ludwig},\ and\ \citenamefont {Lodahl}}]{Uppu20}%
  \BibitemOpen
  \bibfield  {author} {\bibinfo {author} {\bibfnamefont {R.}~\bibnamefont {Uppu}}, \bibinfo {author} {\bibfnamefont {F.~T.}\ \bibnamefont {Pedersen}}, \bibinfo {author} {\bibfnamefont {Y.}~\bibnamefont {Wang}}, \bibinfo {author} {\bibfnamefont {C.~T.}\ \bibnamefont {Olesen}}, \bibinfo {author} {\bibfnamefont {C.}~\bibnamefont {Papon}}, \bibinfo {author} {\bibfnamefont {X.}~\bibnamefont {Zhou}}, \bibinfo {author} {\bibfnamefont {L.}~\bibnamefont {Midolo}}, \bibinfo {author} {\bibfnamefont {S.}~\bibnamefont {Scholz}}, \bibinfo {author} {\bibfnamefont {A.~D.}\ \bibnamefont {Wieck}}, \bibinfo {author} {\bibfnamefont {A.}~\bibnamefont {Ludwig}},\ and\ \bibinfo {author} {\bibfnamefont {P.}~\bibnamefont {Lodahl}},\ }\bibfield  {title} {\bibinfo {title} {Scalable integrated single-photon source},\ }\href {https://doi.org/10.1126/sciadv.abc8268} {\bibfield  {journal} {\bibinfo  {journal} {Science Advances}\ }\textbf {\bibinfo {volume} {6}},\ \bibinfo {pages} {eabc8268} (\bibinfo {year} {2020})}\BibitemShut {NoStop}%
\bibitem [{\citenamefont {Wang}\ \emph {et~al.}(2019)\citenamefont {Wang}, \citenamefont {He}, \citenamefont {Chung}, \citenamefont {Hu}, \citenamefont {Yu}, \citenamefont {Chen}, \citenamefont {Ding}, \citenamefont {Chen}, \citenamefont {Qin}, \citenamefont {Yang}, \citenamefont {Liu}, \citenamefont {Duan}, \citenamefont {Li}, \citenamefont {Gerhardt}, \citenamefont {Winkler}, \citenamefont {Jurkat}, \citenamefont {Wang}, \citenamefont {Gregersen}, \citenamefont {Huo}, \citenamefont {Dai}, \citenamefont {Yu}, \citenamefont {H{\"o}fling}, \citenamefont {Lu},\ and\ \citenamefont {Pan}}]{Wang19}%
  \BibitemOpen
  \bibfield  {author} {\bibinfo {author} {\bibfnamefont {H.}~\bibnamefont {Wang}}, \bibinfo {author} {\bibfnamefont {Y.-M.}\ \bibnamefont {He}}, \bibinfo {author} {\bibfnamefont {T.}~\bibnamefont {Chung}}, \bibinfo {author} {\bibfnamefont {H.}~\bibnamefont {Hu}}, \bibinfo {author} {\bibfnamefont {Y.}~\bibnamefont {Yu}}, \bibinfo {author} {\bibfnamefont {S.}~\bibnamefont {Chen}}, \bibinfo {author} {\bibfnamefont {X.}~\bibnamefont {Ding}}, \bibinfo {author} {\bibfnamefont {M.-C.}\ \bibnamefont {Chen}}, \bibinfo {author} {\bibfnamefont {J.}~\bibnamefont {Qin}}, \bibinfo {author} {\bibfnamefont {X.}~\bibnamefont {Yang}}, \bibinfo {author} {\bibfnamefont {R.-Z.}\ \bibnamefont {Liu}}, \bibinfo {author} {\bibfnamefont {Z.-C.}\ \bibnamefont {Duan}}, \bibinfo {author} {\bibfnamefont {J.-P.}\ \bibnamefont {Li}}, \bibinfo {author} {\bibfnamefont {S.}~\bibnamefont {Gerhardt}}, \bibinfo {author} {\bibfnamefont {K.}~\bibnamefont {Winkler}}, \bibinfo {author} {\bibfnamefont {J.}~\bibnamefont {Jurkat}}, \bibinfo {author}
  {\bibfnamefont {L.-J.}\ \bibnamefont {Wang}}, \bibinfo {author} {\bibfnamefont {N.}~\bibnamefont {Gregersen}}, \bibinfo {author} {\bibfnamefont {Y.-H.}\ \bibnamefont {Huo}}, \bibinfo {author} {\bibfnamefont {Q.}~\bibnamefont {Dai}}, \bibinfo {author} {\bibfnamefont {S.}~\bibnamefont {Yu}}, \bibinfo {author} {\bibfnamefont {S.}~\bibnamefont {H{\"o}fling}}, \bibinfo {author} {\bibfnamefont {C.-Y.}\ \bibnamefont {Lu}},\ and\ \bibinfo {author} {\bibfnamefont {J.-W.}\ \bibnamefont {Pan}},\ }\bibfield  {title} {\bibinfo {title} {Towards optimal single-photon sources from polarized microcavities},\ }\href {https://doi.org/10.1038/s41566-019-0494-3} {\bibfield  {journal} {\bibinfo  {journal} {Nature Photonics}\ }\textbf {\bibinfo {volume} {13}},\ \bibinfo {pages} {770–775} (\bibinfo {year} {2019})}\BibitemShut {NoStop}%
\bibitem [{\citenamefont {Lago-Rivera}\ \emph {et~al.}(2021)\citenamefont {Lago-Rivera}, \citenamefont {Grandi}, \citenamefont {Rakonjac}, \citenamefont {Seri},\ and\ \citenamefont {de~Riedmatten}}]{Lago-Rivera2021}%
  \BibitemOpen
  \bibfield  {author} {\bibinfo {author} {\bibfnamefont {D.}~\bibnamefont {Lago-Rivera}}, \bibinfo {author} {\bibfnamefont {S.}~\bibnamefont {Grandi}}, \bibinfo {author} {\bibfnamefont {J.~V.}\ \bibnamefont {Rakonjac}}, \bibinfo {author} {\bibfnamefont {A.}~\bibnamefont {Seri}},\ and\ \bibinfo {author} {\bibfnamefont {H.}~\bibnamefont {de~Riedmatten}},\ }\bibfield  {title} {\bibinfo {title} {Telecom-heralded entanglement between multimode solid-state quantum memories},\ }\href {https://doi.org/10.1038/s41586-021-03481-8} {\bibfield  {journal} {\bibinfo  {journal} {Nature}\ }\textbf {\bibinfo {volume} {594}},\ \bibinfo {pages} {37} (\bibinfo {year} {2021})}\BibitemShut {NoStop}%
\bibitem [{\citenamefont {Stolk}\ \emph {et~al.}(2024)\citenamefont {Stolk}, \citenamefont {van~der Enden}, \citenamefont {Slater}, \citenamefont {te~Raa-Derckx}, \citenamefont {Botma}, \citenamefont {van Rantwijk}, \citenamefont {Biemond}, \citenamefont {Hagen}, \citenamefont {Herfst}, \citenamefont {Koek}, \citenamefont {Meskers}, \citenamefont {Vollmer}, \citenamefont {van Zwet}, \citenamefont {Markham}, \citenamefont {Edmonds}, \citenamefont {Geus}, \citenamefont {Elsen}, \citenamefont {Jungbluth}, \citenamefont {Haefner}, \citenamefont {Tresp}, \citenamefont {Stuhler}, \citenamefont {Ritter},\ and\ \citenamefont {Hanson}}]{stolk2021}%
  \BibitemOpen
  \bibfield  {author} {\bibinfo {author} {\bibfnamefont {A.~J.}\ \bibnamefont {Stolk}}, \bibinfo {author} {\bibfnamefont {K.~L.}\ \bibnamefont {van~der Enden}}, \bibinfo {author} {\bibfnamefont {M.-C.}\ \bibnamefont {Slater}}, \bibinfo {author} {\bibfnamefont {I.}~\bibnamefont {te~Raa-Derckx}}, \bibinfo {author} {\bibfnamefont {P.}~\bibnamefont {Botma}}, \bibinfo {author} {\bibfnamefont {J.}~\bibnamefont {van Rantwijk}}, \bibinfo {author} {\bibfnamefont {J.~J.~B.}\ \bibnamefont {Biemond}}, \bibinfo {author} {\bibfnamefont {R.~A.~J.}\ \bibnamefont {Hagen}}, \bibinfo {author} {\bibfnamefont {R.~W.}\ \bibnamefont {Herfst}}, \bibinfo {author} {\bibfnamefont {W.~D.}\ \bibnamefont {Koek}}, \bibinfo {author} {\bibfnamefont {A.~J.~H.}\ \bibnamefont {Meskers}}, \bibinfo {author} {\bibfnamefont {R.}~\bibnamefont {Vollmer}}, \bibinfo {author} {\bibfnamefont {E.~J.}\ \bibnamefont {van Zwet}}, \bibinfo {author} {\bibfnamefont {M.}~\bibnamefont {Markham}}, \bibinfo {author} {\bibfnamefont {A.~M.}\ \bibnamefont {Edmonds}},
  \bibinfo {author} {\bibfnamefont {J.~F.}\ \bibnamefont {Geus}}, \bibinfo {author} {\bibfnamefont {F.}~\bibnamefont {Elsen}}, \bibinfo {author} {\bibfnamefont {B.}~\bibnamefont {Jungbluth}}, \bibinfo {author} {\bibfnamefont {C.}~\bibnamefont {Haefner}}, \bibinfo {author} {\bibfnamefont {C.}~\bibnamefont {Tresp}}, \bibinfo {author} {\bibfnamefont {J.}~\bibnamefont {Stuhler}}, \bibinfo {author} {\bibfnamefont {S.}~\bibnamefont {Ritter}},\ and\ \bibinfo {author} {\bibfnamefont {R.}~\bibnamefont {Hanson}},\ }\bibfield  {title} {\bibinfo {title} {Metropolitan-scale heralded entanglement of solid-state qubits},\ }\href {https://doi.org/10.1126/sciadv.adp6442} {\bibfield  {journal} {\bibinfo  {journal} {Science Advances}\ }\textbf {\bibinfo {volume} {10}},\ \bibinfo {pages} {eadp6442} (\bibinfo {year} {2024})},\ \Eprint {https://arxiv.org/abs/https://www.science.org/doi/pdf/10.1126/sciadv.adp6442} {https://www.science.org/doi/pdf/10.1126/sciadv.adp6442} \BibitemShut {NoStop}%
\bibitem [{\citenamefont {Liu}\ \emph {et~al.}(2024)\citenamefont {Liu}, \citenamefont {Luo}, \citenamefont {Yu}, \citenamefont {Wang}, \citenamefont {Wang}, \citenamefont {Hu}, \citenamefont {Li}, \citenamefont {Zheng}, \citenamefont {Yao}, \citenamefont {Yan}, \citenamefont {Teng}, \citenamefont {Jiang}, \citenamefont {Liu}, \citenamefont {Xie}, \citenamefont {Zhang}, \citenamefont {Mao}, \citenamefont {Jiang}, \citenamefont {Zhang}, \citenamefont {Bao},\ and\ \citenamefont {Pan}}]{Liu2024}%
  \BibitemOpen
  \bibfield  {author} {\bibinfo {author} {\bibfnamefont {J.-L.}\ \bibnamefont {Liu}}, \bibinfo {author} {\bibfnamefont {X.-Y.}\ \bibnamefont {Luo}}, \bibinfo {author} {\bibfnamefont {Y.}~\bibnamefont {Yu}}, \bibinfo {author} {\bibfnamefont {C.-Y.}\ \bibnamefont {Wang}}, \bibinfo {author} {\bibfnamefont {B.}~\bibnamefont {Wang}}, \bibinfo {author} {\bibfnamefont {Y.}~\bibnamefont {Hu}}, \bibinfo {author} {\bibfnamefont {J.}~\bibnamefont {Li}}, \bibinfo {author} {\bibfnamefont {M.-Y.}\ \bibnamefont {Zheng}}, \bibinfo {author} {\bibfnamefont {B.}~\bibnamefont {Yao}}, \bibinfo {author} {\bibfnamefont {Z.}~\bibnamefont {Yan}}, \bibinfo {author} {\bibfnamefont {D.}~\bibnamefont {Teng}}, \bibinfo {author} {\bibfnamefont {J.-W.}\ \bibnamefont {Jiang}}, \bibinfo {author} {\bibfnamefont {X.-B.}\ \bibnamefont {Liu}}, \bibinfo {author} {\bibfnamefont {X.-P.}\ \bibnamefont {Xie}}, \bibinfo {author} {\bibfnamefont {J.}~\bibnamefont {Zhang}}, \bibinfo {author} {\bibfnamefont {Q.-H.}\ \bibnamefont {Mao}}, \bibinfo {author}
  {\bibfnamefont {X.}~\bibnamefont {Jiang}}, \bibinfo {author} {\bibfnamefont {Q.}~\bibnamefont {Zhang}}, \bibinfo {author} {\bibfnamefont {X.-H.}\ \bibnamefont {Bao}},\ and\ \bibinfo {author} {\bibfnamefont {J.-W.}\ \bibnamefont {Pan}},\ }\bibfield  {title} {\bibinfo {title} {Creation of memory–memory entanglement in a metropolitan quantum network},\ }\href {https://doi.org/10.1038/s41586-024-07308-0} {\bibfield  {journal} {\bibinfo  {journal} {Nature}\ }\textbf {\bibinfo {volume} {629}},\ \bibinfo {pages} {579} (\bibinfo {year} {2024})}\BibitemShut {NoStop}%
\bibitem [{\citenamefont {Liu}\ \emph {et~al.}(2023)\citenamefont {Liu}, \citenamefont {Zhang}, \citenamefont {Jiang}, \citenamefont {Chen}, \citenamefont {Zhang}, \citenamefont {Pan}, \citenamefont {Ma}, \citenamefont {Dong}, \citenamefont {Xiong}, \citenamefont {Zhang}, \citenamefont {Li}, \citenamefont {Wang}, \citenamefont {Wu}, \citenamefont {Chen}, \citenamefont {You}, \citenamefont {Wang}, \citenamefont {Zhang},\ and\ \citenamefont {Pan}}]{1000kmTwinField}%
  \BibitemOpen
  \bibfield  {author} {\bibinfo {author} {\bibfnamefont {Y.}~\bibnamefont {Liu}}, \bibinfo {author} {\bibfnamefont {W.-J.}\ \bibnamefont {Zhang}}, \bibinfo {author} {\bibfnamefont {C.}~\bibnamefont {Jiang}}, \bibinfo {author} {\bibfnamefont {J.-P.}\ \bibnamefont {Chen}}, \bibinfo {author} {\bibfnamefont {C.}~\bibnamefont {Zhang}}, \bibinfo {author} {\bibfnamefont {W.-X.}\ \bibnamefont {Pan}}, \bibinfo {author} {\bibfnamefont {D.}~\bibnamefont {Ma}}, \bibinfo {author} {\bibfnamefont {H.}~\bibnamefont {Dong}}, \bibinfo {author} {\bibfnamefont {J.-M.}\ \bibnamefont {Xiong}}, \bibinfo {author} {\bibfnamefont {C.-J.}\ \bibnamefont {Zhang}}, \bibinfo {author} {\bibfnamefont {H.}~\bibnamefont {Li}}, \bibinfo {author} {\bibfnamefont {R.-C.}\ \bibnamefont {Wang}}, \bibinfo {author} {\bibfnamefont {J.}~\bibnamefont {Wu}}, \bibinfo {author} {\bibfnamefont {T.-Y.}\ \bibnamefont {Chen}}, \bibinfo {author} {\bibfnamefont {L.}~\bibnamefont {You}}, \bibinfo {author} {\bibfnamefont {X.-B.}\ \bibnamefont {Wang}}, \bibinfo
  {author} {\bibfnamefont {Q.}~\bibnamefont {Zhang}},\ and\ \bibinfo {author} {\bibfnamefont {J.-W.}\ \bibnamefont {Pan}},\ }\bibfield  {title} {\bibinfo {title} {Experimental twin-field quantum key distribution over 1000 km fiber distance},\ }\href {https://doi.org/10.1103/PhysRevLett.130.210801} {\bibfield  {journal} {\bibinfo  {journal} {Phys. Rev. Lett.}\ }\textbf {\bibinfo {volume} {130}},\ \bibinfo {pages} {210801} (\bibinfo {year} {2023})}\BibitemShut {NoStop}%
\bibitem [{\citenamefont {Caspar}\ \emph {et~al.}(2022)\citenamefont {Caspar}, \citenamefont {Oudot}, \citenamefont {Sekatski}, \citenamefont {Maring}, \citenamefont {Martin}, \citenamefont {Sangouard}, \citenamefont {Zbinden},\ and\ \citenamefont {Thew}}]{Caspar2022}%
  \BibitemOpen
  \bibfield  {author} {\bibinfo {author} {\bibfnamefont {P.}~\bibnamefont {Caspar}}, \bibinfo {author} {\bibfnamefont {E.}~\bibnamefont {Oudot}}, \bibinfo {author} {\bibfnamefont {P.}~\bibnamefont {Sekatski}}, \bibinfo {author} {\bibfnamefont {N.}~\bibnamefont {Maring}}, \bibinfo {author} {\bibfnamefont {A.}~\bibnamefont {Martin}}, \bibinfo {author} {\bibfnamefont {N.}~\bibnamefont {Sangouard}}, \bibinfo {author} {\bibfnamefont {H.}~\bibnamefont {Zbinden}},\ and\ \bibinfo {author} {\bibfnamefont {R.}~\bibnamefont {Thew}},\ }\bibfield  {title} {\bibinfo {title} {Local and scalable detection of genuine multipartite single-photon path entanglement},\ }\href {https://doi.org/10.22331/q-2022-03-22-671} {\bibfield  {journal} {\bibinfo  {journal} {{Quantum}}\ }\textbf {\bibinfo {volume} {6}},\ \bibinfo {pages} {671} (\bibinfo {year} {2022})}\BibitemShut {NoStop}%
\bibitem [{\citenamefont {Monteiro}\ \emph {et~al.}(2015)\citenamefont {Monteiro}, \citenamefont {Vivoli}, \citenamefont {Guerreiro}, \citenamefont {Martin}, \citenamefont {Bancal}, \citenamefont {Zbinden}, \citenamefont {Thew},\ and\ \citenamefont {Sangouard}}]{Monteiro}%
  \BibitemOpen
  \bibfield  {author} {\bibinfo {author} {\bibfnamefont {F.}~\bibnamefont {Monteiro}}, \bibinfo {author} {\bibfnamefont {V.~C.}\ \bibnamefont {Vivoli}}, \bibinfo {author} {\bibfnamefont {T.}~\bibnamefont {Guerreiro}}, \bibinfo {author} {\bibfnamefont {A.}~\bibnamefont {Martin}}, \bibinfo {author} {\bibfnamefont {J.-D.}\ \bibnamefont {Bancal}}, \bibinfo {author} {\bibfnamefont {H.}~\bibnamefont {Zbinden}}, \bibinfo {author} {\bibfnamefont {R.~T.}\ \bibnamefont {Thew}},\ and\ \bibinfo {author} {\bibfnamefont {N.}~\bibnamefont {Sangouard}},\ }\bibfield  {title} {\bibinfo {title} {Revealing genuine optical-path entanglement},\ }\href {https://doi.org/10.1103/PhysRevLett.114.170504} {\bibfield  {journal} {\bibinfo  {journal} {Phys. Rev. Lett.}\ }\textbf {\bibinfo {volume} {114}},\ \bibinfo {pages} {170504} (\bibinfo {year} {2015})}\BibitemShut {NoStop}%
\bibitem [{\citenamefont {Andersen}\ \emph {et~al.}(2016)\citenamefont {Andersen}, \citenamefont {Gehring}, \citenamefont {Marquardt},\ and\ \citenamefont {Leuchs}}]{Andersen30YearsSqueezing}%
  \BibitemOpen
  \bibfield  {author} {\bibinfo {author} {\bibfnamefont {U.~L.}\ \bibnamefont {Andersen}}, \bibinfo {author} {\bibfnamefont {T.}~\bibnamefont {Gehring}}, \bibinfo {author} {\bibfnamefont {C.}~\bibnamefont {Marquardt}},\ and\ \bibinfo {author} {\bibfnamefont {G.}~\bibnamefont {Leuchs}},\ }\bibfield  {title} {\bibinfo {title} {30 years of squeezed light generation},\ }\href {https://doi.org/10.1088/0031-8949/91/5/053001} {\bibfield  {journal} {\bibinfo  {journal} {Physica Scripta}\ }\textbf {\bibinfo {volume} {91}},\ \bibinfo {pages} {053001} (\bibinfo {year} {2016})}\BibitemShut {NoStop}%
\bibitem [{\citenamefont {Steffinlongo}(2024)}]{git}%
  \BibitemOpen
  \bibfield  {author} {\bibinfo {author} {\bibfnamefont {A.}~\bibnamefont {Steffinlongo}},\ }\href {https://github.com/AnnaSteffinlongo/DIQKD-with-single-photons} {\bibinfo {title} {https://github.com/annasteffinlongo/diqkd-with-single-photons}} (\bibinfo {year} {2024})\BibitemShut {NoStop}%
\bibitem [{\citenamefont {Vivoli}\ \emph {et~al.}(2015)\citenamefont {Vivoli}, \citenamefont {Sekatski}, \citenamefont {Bancal}, \citenamefont {Lim}, \citenamefont {Martin}, \citenamefont {Thew}, \citenamefont {Hugo.Zbinden}, \citenamefont {Nicolas.Gisin},\ and\ \citenamefont {Nicola.Sangouard}}]{CapraraVivoli2015}%
  \BibitemOpen
  \bibfield  {author} {\bibinfo {author} {\bibfnamefont {V.~C.}\ \bibnamefont {Vivoli}}, \bibinfo {author} {\bibfnamefont {P.}~\bibnamefont {Sekatski}}, \bibinfo {author} {\bibfnamefont {J.}~\bibnamefont {Bancal}}, \bibinfo {author} {\bibfnamefont {C.~W.}\ \bibnamefont {Lim}}, \bibinfo {author} {\bibfnamefont {A.}~\bibnamefont {Martin}}, \bibinfo {author} {\bibfnamefont {R.}~\bibnamefont {Thew}}, \bibinfo {author} {\bibnamefont {Hugo.Zbinden}}, \bibinfo {author} {\bibnamefont {Nicolas.Gisin}},\ and\ \bibinfo {author} {\bibnamefont {Nicola.Sangouard}},\ }\bibfield  {title} {\bibinfo {title} {Comparing different approaches for generating random numbers device‐independently using a photon pair source},\ }\href {https://doi.org/10.1088/1367-2630/17/2/023023} {\bibfield  {journal} {\bibinfo  {journal} {New Journal of Physics}\ }\textbf {\bibinfo {volume} {17}},\ \bibinfo {pages} {023023} (\bibinfo {year} {2015})}\BibitemShut {NoStop}%
\bibitem [{\citenamefont {Liu}\ \emph {et~al.}(2022)\citenamefont {Liu}, \citenamefont {Zhang}, \citenamefont {Zhen}, \citenamefont {Li}, \citenamefont {Liu}, \citenamefont {Fan}, \citenamefont {Xu}, \citenamefont {Zhang},\ and\ \citenamefont {Pan}}]{SPDC}%
  \BibitemOpen
  \bibfield  {author} {\bibinfo {author} {\bibfnamefont {W.-Z.}\ \bibnamefont {Liu}}, \bibinfo {author} {\bibfnamefont {Y.-Z.}\ \bibnamefont {Zhang}}, \bibinfo {author} {\bibfnamefont {Y.-Z.}\ \bibnamefont {Zhen}}, \bibinfo {author} {\bibfnamefont {M.-H.}\ \bibnamefont {Li}}, \bibinfo {author} {\bibfnamefont {Y.}~\bibnamefont {Liu}}, \bibinfo {author} {\bibfnamefont {J.}~\bibnamefont {Fan}}, \bibinfo {author} {\bibfnamefont {F.}~\bibnamefont {Xu}}, \bibinfo {author} {\bibfnamefont {Q.}~\bibnamefont {Zhang}},\ and\ \bibinfo {author} {\bibfnamefont {J.-W.}\ \bibnamefont {Pan}},\ }\bibfield  {title} {\bibinfo {title} {Toward a photonic demonstration of device-independent quantum key distribution},\ }\href {https://doi.org/10.1103/PhysRevLett.129.050502} {\bibfield  {journal} {\bibinfo  {journal} {Phys. Rev. Lett.}\ }\textbf {\bibinfo {volume} {129}},\ \bibinfo {pages} {050502} (\bibinfo {year} {2022})}\BibitemShut {NoStop}%
\bibitem [{\citenamefont {Christ}\ \emph {et~al.}(2011)\citenamefont {Christ}, \citenamefont {Brecht}, \citenamefont {Mauerer},\ and\ \citenamefont {Silberhorn}}]{Christ2011}%
  \BibitemOpen
  \bibfield  {author} {\bibinfo {author} {\bibfnamefont {A.}~\bibnamefont {Christ}}, \bibinfo {author} {\bibfnamefont {B.}~\bibnamefont {Brecht}}, \bibinfo {author} {\bibfnamefont {W.}~\bibnamefont {Mauerer}},\ and\ \bibinfo {author} {\bibfnamefont {C.}~\bibnamefont {Silberhorn}},\ }\bibfield  {title} {\bibinfo {title} {Optimized generation of heralded fock states using parametric down-conversion},\ }\href {https://doi.org/10.1088/1367-2630/13/3/033027} {\bibfield  {journal} {\bibinfo  {journal} {New Journal of Physics}\ }\textbf {\bibinfo {volume} {13}},\ \bibinfo {pages} {033027} (\bibinfo {year} {2011})}\BibitemShut {NoStop}%
\bibitem [{\citenamefont {Muljarov}\ and\ \citenamefont {Zimmermann}(2004)}]{Muljarov2004}%
  \BibitemOpen
  \bibfield  {author} {\bibinfo {author} {\bibfnamefont {E.~A.}\ \bibnamefont {Muljarov}}\ and\ \bibinfo {author} {\bibfnamefont {R.}~\bibnamefont {Zimmermann}},\ }\bibfield  {title} {\bibinfo {title} {Dephasing in quantum dots: Quadratic coupling to acoustic phonons},\ }\href {https://doi.org/10.1103/PhysRevLett.93.237401} {\bibfield  {journal} {\bibinfo  {journal} {Phys. Rev. Lett.}\ }\textbf {\bibinfo {volume} {93}},\ \bibinfo {pages} {237401} (\bibinfo {year} {2004})}\BibitemShut {NoStop}%
\bibitem [{\citenamefont {Zhao}\ \emph {et~al.}(2020)\citenamefont {Zhao}, \citenamefont {Ma}, \citenamefont {R\"using},\ and\ \citenamefont {Mookherjea}}]{Zhao2020}%
  \BibitemOpen
  \bibfield  {author} {\bibinfo {author} {\bibfnamefont {J.}~\bibnamefont {Zhao}}, \bibinfo {author} {\bibfnamefont {C.}~\bibnamefont {Ma}}, \bibinfo {author} {\bibfnamefont {M.}~\bibnamefont {R\"using}},\ and\ \bibinfo {author} {\bibfnamefont {S.}~\bibnamefont {Mookherjea}},\ }\bibfield  {title} {\bibinfo {title} {High quality entangled photon pair generation in periodically poled thin-film lithium niobate waveguides},\ }\href {https://doi.org/10.1103/PhysRevLett.124.163603} {\bibfield  {journal} {\bibinfo  {journal} {Phys. Rev. Lett.}\ }\textbf {\bibinfo {volume} {124}},\ \bibinfo {pages} {163603} (\bibinfo {year} {2020})}\BibitemShut {NoStop}%
\bibitem [{\citenamefont {Gonz\'alez-Ruiz}\ \emph {et~al.}(2022)\citenamefont {Gonz\'alez-Ruiz}, \citenamefont {Das}, \citenamefont {Lodahl},\ and\ \citenamefont {S\o{}rensen}}]{Gonzalez2022}%
  \BibitemOpen
  \bibfield  {author} {\bibinfo {author} {\bibfnamefont {E.~M.}\ \bibnamefont {Gonz\'alez-Ruiz}}, \bibinfo {author} {\bibfnamefont {S.~K.}\ \bibnamefont {Das}}, \bibinfo {author} {\bibfnamefont {P.}~\bibnamefont {Lodahl}},\ and\ \bibinfo {author} {\bibfnamefont {A.~S.}\ \bibnamefont {S\o{}rensen}},\ }\bibfield  {title} {\bibinfo {title} {Violation of bell's inequality with quantum-dot single-photon sources},\ }\href {https://doi.org/10.1103/PhysRevA.106.012222} {\bibfield  {journal} {\bibinfo  {journal} {Phys. Rev. A}\ }\textbf {\bibinfo {volume} {106}},\ \bibinfo {pages} {012222} (\bibinfo {year} {2022})}\BibitemShut {NoStop}%
\end{thebibliography}%

\end{document}


\allowdisplaybreaks

\clearpage
\appendix

\begin{center}
\large{\textbf{\textsc{Supplementary Material}}}

\end{center}

\section{Squeezed Displaced Measurements} \label{App:measurements}
In this section, we present the novel measurements performed by Alice and Bob to observe a Bell inequality violation. In particular, we focus on a 2-outcome POVM comprising a displacement and a squeezing operator, followed by a single-photon detector. The two possible outcomes we consider are -1 when the detector does not click, and +1 when it clicks. The corresponding POVM elements read 
\begin{align}
    \Pi_{-1}^{(\xi,\alpha)}&=D(\alpha)S(\xi)\ketbra{0}{0} S^\dagger(\xi)D^\dagger(\alpha) , \label{eq: POVM  click}\\
    \Pi_{+1}^{(\xi,\alpha)}& =\mathds{1}-\Pi_{-1}^{(\xi,\alpha)}.\label{eq: POVM no click}
\end{align}
Here $D(\alpha)=e^{\alpha^*\hat{a}-\alpha\hat{a}^{\dagger}}$ and $S(\xi)=e^{\frac{1}{2}(\hat{a}^2\xi^*-\hat{a}^{\dagger 2}\xi)}$ correspond to the displacement and squeezing operator, respectively. 
The arguments of these operators are complex numbers, such as $\xi = |\xi| e^{i \phi}$ and $\alpha = |\alpha| e^{i \theta}$. 
Since we are interested in using such measurement on the heralded state in Eq.\eqrefheraldedstate of the main text, we focus on its projection on the qubit space spanned by $\ket{0}$ and $\ket{1}$. Thus, an arbitrary 2-outcome qubit POVM can be written as \cite{CapraraVivoli2015}
\begin{flalign}\label{eq:generalPOVM}
    E_1 &= \mu_{\ket{\vec{n}}} \ket{\vec{n}}\bra{\vec{n}} + (1-\mu_{\ket{\vec{n}}}) r_1 \mathds{1},\\
    E_2&= \mu_{\ket{\vec{n}}} \ket{-\vec{n}}\bra{-\vec{n}} + (1-\mu_{\ket{\vec{n}}}) r_2 \mathds{1}
\end{flalign} 

 where $\mu_{\ket{\vec{n}}} \in [0,1]$ and $r_1=1-r_2 \in [0,1]$ quantify the projective and the random part of the measurement, respectively (random in the sense that it gives the same value for all the incoming states). The two vectors $\ket{\pm \vec{n}}$ are orthogonal and $\mu_{\ket{\vec{n}}}$ quantifies the ability of the POVM to distinguish between them. To identify the projective part of the POVM $\{ \Pi_{+1}^{(\xi,\alpha)}, \Pi_{-1}^{(\xi,\alpha)} \}$ when projected into the $\{\ket 0,\ket 1\}$ subspace, we first write a displaced squeezed vacuum in the Fock basis:  $D(\alpha)S(\xi)\ket{0}=\sum_n c_n \ket{n}.$ 
Then, the matrix elements of the POVM \eqref{eq: POVM click} are  
\begin{align}
    \Pi_{-1}^{(\xi,\alpha)}(n,m)=c_n c_m^*, \label{eq: measurement Alice/Bob}
\end{align}
where we explicitly write
\begin{equation}
    c_k = \frac{e^{\frac{|\alpha|^2 }{2} (  e^{i (\phi+2\theta)}\tanh{(|\xi|)} -1 )}}{k! \cosh{(|\xi|)}} \left(\sqrt{\frac{ e^{-i \phi}}{2} \tanh{(|\xi|)}}\right)^k H_k \left(\frac{|\alpha| e^{i \theta}}{\sqrt{e^{i \phi} \sinh{(2|\xi|)}}}\right),
\end{equation}
 
being $H_k(x)$ the Hermite polynomials of order $k$. At this stage, we have a close form of the POVM in the Fock basis for arbitrary dimensions. For example, if we restrict ourselves to a qubit space spanned by 0 and 1 photons, we get the POVM element 
    \begin{align}\label{eq:measurements one photon}
            \Pi_{-1} = \begin{pmatrix}
    	   \vartheta \sech{(|\xi|)} & |\alpha| \vartheta e^{-i \theta}\sech{(|\xi|)}^2\\
    		|\alpha| \vartheta e^{i \theta}\sech{(|\xi|)}^2 & |\alpha|^2 \vartheta \sech{(|\xi|)}^3 
    	\end{pmatrix}, 
    \end{align}

     with  $ \vartheta = \exp{|\alpha|^2(\cos{(\phi - 2\theta)} \tanh{(|\xi|)} - 1)} $. 
     
     We now aim to compute the projective part of the POVM $\mu_{\ket{\vec{n}}}$ in the direction $\ket{\vec{n}}$, following Eq.~\eqref{eq:generalPOVM}, which is given by   
\begin{equation}
\label{eq: projective part}
\mu_{\ket{\vec{n}}}=\text{Tr}((E_1-E_2)\sigma^{\ket{\vec{n}}}),
\end{equation}
 
where $\sigma^{\ket{\vec{n}}}=\ket{\vec{n}}\bra{\vec{n}}-\ket{-\vec{n}}\bra{-\vec{n}}$. Inserting the POVM $\{\Pi_{+1}^{(\xi,\alpha)},\Pi_{-1}^{(\xi,\alpha)}\}$ in Eq.~\eqref{eq: projective part} we get
 
 \begin{equation}
 \mu_{\ket{+\vec{n}}}(\xi,\alpha)=\text{Tr}((\Pi_{+1}^{(\xi,\alpha)}-\Pi_{-1}^{(\xi,\alpha)})\sigma^{\ket{\vec{n}}}).
 \end{equation}
 
 This quantifies how well we can mimic a Pauli measurement in the direction $\ket{+\vec{n}}$ defined in the photonic qubit space using the POVM described in Fig.\refmeasurements in the main text.
 For each $\ket{+\vec{n}}$ we then optimize $\mu_{\ket{+\vec{n}}} (\xi,\alpha)$ according to 
 \begin{equation}
\mu_{\ket{+\vec{n}}}^{\text{max}}=\max_{\xi,\alpha}\mu_{\ket{+\vec{n}}} (\xi,\alpha).
 \end{equation}

\begin{figure}[htbp] 
        \centering
        \includegraphics[width=0.6\textwidth]{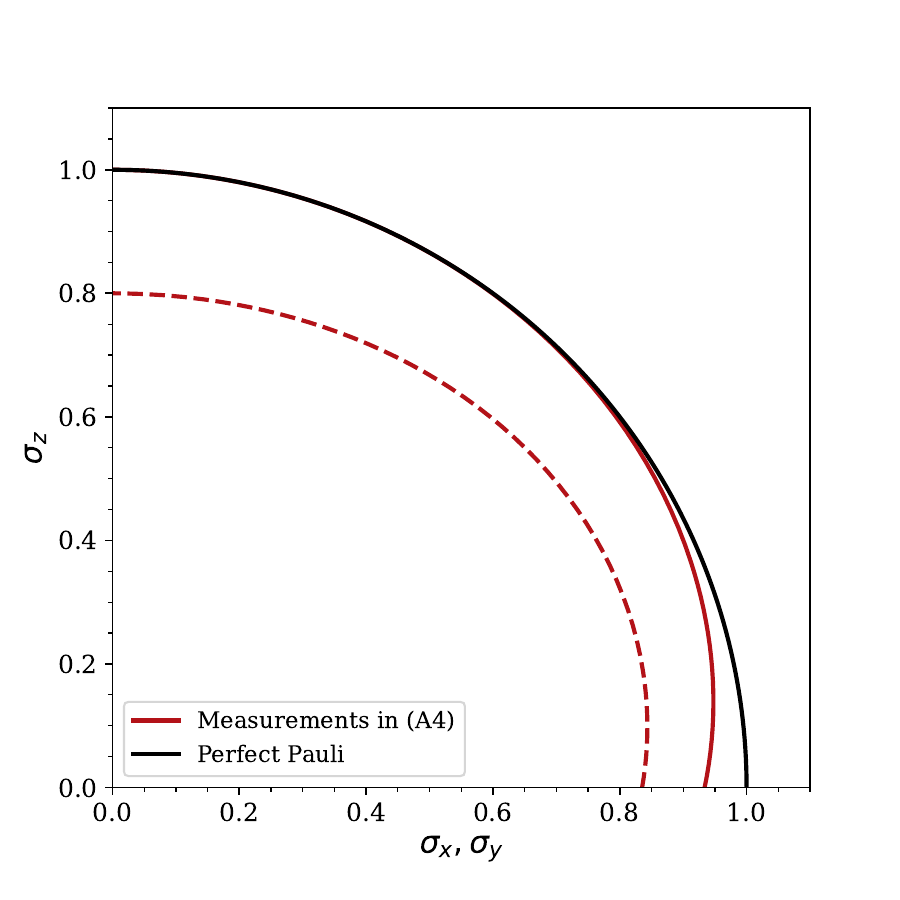}
        \caption{Comparison of the projection $\mu^\text{max}_{\ket{+\vec{n}}}$ along a given direction $\ket{\vec{n}}$ of POVM with described in \eqref{eq: measurement Alice/Bob} (red) and using noiseless Pauli matrices (black). Solid lines indicate ideal measurements with local efficiency $\tilde{\eta}_L = 100\%$, while dashed lines represent measurements with $\tilde{\eta}_L = 80\%$, both with detection efficiency $\tilde{\eta}_D = 100\%$. Interestingly, for imperfect measurements, the projection along $\sigma_{x,y}$ is larger than that along $\sigma_z$.} \label{fig:projections}
    \end{figure}

In Fig.~\ref{fig:projections} we present the projection $\mu_{\ket{+\vec{n}}}^{\text{max}}$ along all the possible directions in the positive $\sigma_x\sigma_z$ and $\sigma_y\sigma_z$ planes, where ${\sigma_x,\sigma_y,\sigma_z}$ are the Pauli matrices. 
 

     So far, we have considered measurement with perfect efficiency. To account for losses occurring before the squeezing operation, characterized by the non-unit efficiency $\tilde{\eta}_L$, we apply a loss channel to the state and use the POVM expression given in Eq.~\eqref{eq: measurement Alice/Bob}. Note that the detector efficiency $\tilde{\eta}_D$ has to be treated separately since a loss channel does not commute with a squeezing operation. 
     When considering non-ideal detection efficiency, the POVM element corresponding to a no-click event of a single photon detector with efficiency $\tilde{\eta}_D$ is given by $ (1-\tilde{\eta}_D)^{\hat{a}^\dagger \hat{a}}.$ Applying displacement and squeezing, the POVM elements become
\begin{align}
\label{eq:noisy POVM}
    \Pi_{-1}^{(\xi,\alpha,\tilde{\eta}_D)} &= D(\alpha) \,S(\xi)\, (1-\tilde{\eta}_D)^{\hat{a}^\dagger \hat{a}} \, S^\dagger(\xi)\, D^\dagger(\alpha) , \\
    \Pi_{+1}^{(\xi,\alpha,\tilde{\eta}_D)}& =\mathds{1}-\Pi_{-1}^{(\xi,\alpha,\tilde{\eta}_D)}.\label{eq: POVM no click}
\end{align}
We compute these POVM elements numerically (see \cite{git}) fixing a cutoff of the Hilbert space equal to 5. Throughout this work, we restrict $|\alpha|<1$ and $|\xi|<1$ to ensure that the matrix elements in \eqref{eq:noisy POVM} decay rapidly with photon number, allowing for accurate numerical simulation as the dimension increases.




\section{Heralded State with Ideal Source}\label{app:stateT}

In this section, we derive the heralded state shared by Alice and Bob, based on the setup illustrated in Fig.\refsetup. For simplicity and to provide a clear and intuitive understanding of the scheme, we begin with the ideal case in which the source emits a pure two-mode single-photon state, $\ket{\psi_0} = \ket{11}_{AB}$. While the realistic implementation is addressed in the next section, the analysis here serves to illustrate the basic functioning of the scheme in an ideal scenario. 
According to the setup in Fig.\refsetup, the photons are sent into an unbalanced beamsplitter with transmittance $T$, which transforms the state as follows: 
 
        \begin{align}
            \ket{1} &\rightarrow  \sqrt{T} \ket{01} + \sqrt{1-T} \ket{10}. \label{eq:BS transformation one photon}
        \end{align}
 
     Applying this transformation to the initial state $\ket{\psi_0} $, we can construct the density matrix $\rho_1$ after the unbalanced beamsplitter. 
     The resulting density matrix is given by $\rho_1=\ketbra{\psi_1}$, where:

  \begin{align}
        \ket{\psi_1} = \big(T \ket{0011} + \sqrt{T(1-T)}\left(\ket{0110}+ \ket{1001}\right)+ (1-T)\ket{1100}\big)_{ABC_1C_2} . \label{eq: rho 1}
    \end{align}

    In this expression, the first ($A$) and second ($B$) path modes correspond to the reflected photons from Alice's and Bob's unbalanced beamsplitters, respectively. The third ($C_1$) and fourth ($C_2$) path modes represent the transmitted photons. 
    Before these photons in the $C$ modes reach the central station, they pass through a lossy channel, which we model as a beamsplitter where the reflected photons are sent to the environment. Note that Charlie's balanced beamsplitter commutes with the beamsplitter used to model the detector efficiency $\eta_D$. As a result, the effects of channel losses and detector imperfections can be combined into a single effective parameter. This allows us to absorb Charlie’s beamsplitter into the detection model, leading to a total heralding efficiency given by $\eta_H = \eta_D\sqrt{\eta_C}$,where $\eta_C$ represents the transmission efficiency of the channel connecting Alice and Bob to Charlie.
    Thus, when we have an exclusively one-photon contribution (e.g., terms proportional to $T$) the channel transforms as follows:
 
    \begin{align}
        \rho \rightarrow \eta_H\rho + (1-\eta_H)\ketbra{00}{00},
    \end{align}
 
    For terms proportional to $\sqrt{T^3}$, the channel behaves as:
 
    \begin{align}
        \ketbra{11}{01} \rightarrow \sqrt{\eta_H}(\eta_H  \ketbra{11}{01} + (1-\eta_H) \ketbra{10}{00} ), \\
        \ketbra{10}{11} \rightarrow \sqrt{\eta_H}(\eta_H  \ketbra{10}{11} + (1-\eta_H) \ketbra{00}{01} ),
    \end{align}
 
    and similarly for the two other terms. In the particular case of the state $\ket{11}_{C_1 C_2}$ the channel acts as:
 
    \begin{align}
        \ketbra{11}{11} \rightarrow \eta_H^2\ketbra{11}{11} + \eta_H(1-\eta_H)(\ketbra{10}{10}+ \ketbra{01}{01}) + (1-\eta_H)^2 \ketbra{00}{00}.
    \end{align}
 
    Applying these transformations to the state \eqref{eq: rho 1} we can construct the density matrix that describes the state reaching the central station. At this point, Charlie performs the following projective measurement:
 
    \begin{align}
        \hat{\Pi}_{C_1C_2} = \sum^\infty_{n=1} \text{BS} ( \ket{n 0}\bra{n 0}  )\text{BS}^\dagger,
    \end{align}
    with BS being the unitary matrix of a $50/50$ beamsplitter.
    Note that this implies that only the events heralded by one of the two detectors at Charlie's station are considered, while the other half are discarded. Indeed, the state depends on which detector clicks: to leading order in $T$, Charlie heralds the state $(\ket{01} \pm \ket{10})/\sqrt{2}$, with the sign determined by the detector that clicks. In scenarios where Alice and Bob perform ideal Pauli measurements, they obtain the same CHSH value for both heralded states by measuring the same observables and appropriately relabeling their inputs based on Charlie’s outcome. However, this symmetry no longer holds for the imperfect measurements implemented in our scheme. As described in Eq.~\eqref{eq:generalPOVM}, our measurement can be modeled as a mixture of a shrunken projective (Pauli) measurement and white noise, represented by the identity operator. This added noise component breaks the symmetry between the two heralded states, making it impossible to retrieve the same CHSH value through input relabeling. For this reason, we restrict our analysis to a single heralding event. That said, we obtain the heralded state
    \begin{align}
        \rho_H =  \frac{1}{2}(1-T)(\ketbra{01}{01}+\ketbra{01}{10}+\ketbra{10}{01}+ \ketbra{10}{10})+T\ketbra{00}{00},\label{eq:heralded state one photon no aproximation }
    \end{align}
    with probability $P_H=  T\eta_H$, under the assumption $\eta_H^2 \ll \eta_H$. It is important to highlight that by taking the unnormalized heralded state and approximating $T \rightarrow 0$, we can recover the entangled state\eqrefheraldedstate. 
    Once the state in Eq.~\eqref{eq:heralded state one photon no aproximation } is heralded, it passes through a channel with efficiency $\tilde{\eta}_L$. Consequently, the state measured by Alice and Bob becomes
 
    \begin{align}\label{eq:stateT}
        \bar{\rho}_H = \frac{1}{2}\left[\tilde{\eta}_L(1-T)(\ketbra{01}{01}+\ketbra{01}{10}+\ketbra{10}{01}+ \ketbra{10}{10})+2(1-\tilde{\eta}_L + T \tilde{\eta}_L)\ketbra{00}{00}\right].
    \end{align}

\section{Analysis of the Protocol with Realistic Sources}\label{App:realistic_source}
This section analyzes the impact of using realistic single-photon sources on the heralded state and the resulting key rate. We begin by modeling two widely used sources—spontaneous parametric down-conversion (SPDC) and quantum dots (QDs)—and compare their performance. We then examine how the key rate depends on imperfections in the source parameters, local efficiencies and the transittance parameter $T$.

\subsection{Realistic Single-Photon Sources}
We consider two leading technologies for single-photon generation: SPDC and QDs. These platforms differ significantly in their physical principles and operational characteristics. SPDC sources typically achieve generation rates in the MHz range \cite{SPDC}, whereas QDs can operate at repetition rates up to hundreds of MHz while offering high photon indistinguishability \cite{Tomm_2021}.
To enable a fair comparison, we model both sources using two key parameters: the Hong-Ou-Mandel (HOM) visibility $V$, which quantifies photon indistinguishability \cite{Gonzalez2022}, and the second-order correlation function $g^{(2)}$, which captures the probability of multi-photon emission. For SPDC sources, imperfections in $V$ typically arise from spectral and temporal mode mismatch due to properties of the nonlinear crystal \cite{Christ2011}. In the case of QDs, the main limitation is dephasing caused by phonon interactions \cite{Muljarov2004}.

To capture these imperfections in a unified framework and analyze their impact, we now formalize the underlying photonic states and their spectral structure. We begin by denoting a single-mode single photon as $\ket{1} = a^{\dagger} \ket{0}$. In practice, such a state has a frequency and time distribution, which becomes relevant when multiple photons are involved in the process. We denote the spectral density of Alice's photon as \( f(\omega) \) and that of Bob's photon as \( g(\omega) \). AAlice and Bob each have access to two spatial modes. The first are labeled $A_1$ and $B_1$, respectively, and the second correspond to modes sent to Charlie, labeled $A_2=C_1$ and $B_2=C_2$. Alice’s operator associated with mode $A_1$ is defined as: 
\[
a_1 = \int_{-\infty}^{\infty} f_1(\omega) \hat{a}_1(\omega) \, d\omega,
\]
where \( \hat{a}_1(\omega) \) is the photon annihilation operator at frequency \( \omega \), satisfying the continuous frequency commutation relations:
\[
[\hat{a}_1(\omega), \hat{a}_1^\dagger(\omega')] = \delta(\omega - \omega'), \quad [\hat{a}_1(\omega), \hat{a}_1(\omega')] = [\hat{a}_1^\dagger(\omega), \hat{a}_1^\dagger(\omega')] = 0.
\]
Analogous operators $b_1$, $c_1$ and $c_2$ can be defined for Bob and Charlie.
The functions \( f \) and \( g \) are normalized, and we denote their overlap by \( \alpha \):
\begin{equation}
    \label{Eq:Overlap}
\int_{-\infty}^{\infty} d\omega \, f(\omega) f^*(\omega) = 1, \quad \int_{-\infty}^{\infty} d\omega \, f(\omega) g^*(\omega) = \alpha.
\end{equation}
We define \( V = \alpha \alpha^* \) as the visibility between the two photons, corresponding to the visibility of interference in a Hong-Ou-Mandel experiment. Before the heralding event, the state component contributing to successful detection is: 
\[
\ket{\psi}=\frac{1}{\sqrt{2}}(\ket{0110}+\ket{1001})_{ABC_1C_2}.
\]
When accounting for different spectral densities for Alice and Bob, this state becomes:
\[
\ket{\psi}_{ABC_1C_2} = \frac{1}{\sqrt{2}} \int_{-\infty}^{\infty} d\omega_1 d\omega_2 f(\omega_1) g(\omega_2) \left( a_1^\dagger(\omega_1) c_2^\dagger(\omega_2) + b_1^\dagger(\omega_2) c_1^\dagger(\omega_1) \right) \ket{\bar{0}}.
\]
For detection at Charlie's station, we assume a flat frequency distribution. The POVM element corresponding to a click in one detector and no click in the other is given by:
\[
\Pi = \int_{-\infty}^{\infty} d\omega \, c_1^\dagger(\omega) \ket{00}\bra{00}_{C_1C_2} c_1(\omega),
\]
which, after Charlie's beam splitter, can be written as:
\[
\Pi_{BS} = \frac{1}{2} \int_{-\infty}^{\infty} d\omega \, \left( c_1^\dagger(\omega) + c_2^\dagger(\omega) \right) \ket{00}\bra{00}_{C_1C_2} \left( c_1(\omega) + c_2(\omega) \right).
\]
We now compute the state after a click event at Charlie’s station:
\[
\rho_H = \frac{1}{N} \mathrm{Tr}_C \left( \Pi_{BS} \rho \Pi_{BS} \right),
\]
where $\rho=\ketbra{\psi}{\psi}_{ABC_1C_2}$ and $N$ is a normalization factor.
In particular, applying $\Pi_{BS}$ 
to the state $\ket{\psi}_{ABC_1C_2}$
yields: 
\[
\Pi_{BS}\ket{\psi}_{ABC_1C_2} =\int_{-\infty}^{\infty} d\omega d\omega_1 d\omega_2  \frac{1}{2\sqrt{2}}\left( c_1^\dagger(\omega) + c_2^\dagger(\omega) \right) \ket{00}_{C_1C_2}f(\omega_1) g(\omega_2)[\delta(\omega-\omega_2)a_1^\dagger(\omega_1)+\delta(\omega-\omega_1))b_1^\dagger(\omega_2)]\ket{00}_{AB}.
\]

After tracing out Charlie's subsystem and integrating the delta function, we get

\begin{eqnarray}
\nonumber    
\rho_H =\frac{1}{2}(&\int_{-\infty}^{\infty} d\omega_1^{'} d\omega_1 d\omega_2  f(\omega_1) g(\omega_2)g^{*}(\omega_2)a_1^\dagger(\omega_1)\ket{00}\bra{00}a_1(\omega_1^{'})f^{*}(\omega_1^{'})\\
\nonumber  
+& \int_{-\infty}^{\infty} d\omega_2^{'} d\omega_2 d\omega_1  g(\omega_2) f(\omega_1)f^{*}(\omega_1)b_1^\dagger(\omega_2)\ket{00}\bra{00}b_1(\omega_2^{'})g^{*}(\omega_2^{'})  \\
\nonumber  
+&\int_{-\infty}^{\infty} d\omega_1^{'} d\omega_2 d\omega  g(\omega_2) f(\omega)g^{*}(\omega)b_1^\dagger(\omega_2)\ket{00}\bra{00}a_1(\omega_1^{'})f^{*}(\omega_1^{'}) \\
+& \int_{-\infty}^{\infty} d\omega_2^{'} d\omega d\omega_1  f(\omega_1) g(\omega)f^{*}(\omega)a_1^\dagger(\omega_1)\ket{00}\bra{00}b_1(\omega_2^{'})g^{*}(\omega_2^{'})).  
\end{eqnarray}

Using $\int_{-\infty}^{\infty} d\omega a_1^\dagger(\omega)\ket{00}=\ket{10},$ and $\int_{-\infty}^{\infty} d\omega b_1^\dagger(\omega_2)\ket{00}=\ket{01}$, together with Eq.
\eqref{Eq:Overlap}, we end up with:
\begin{eqnarray}
\nonumber    
\rho_H =\frac{1}{2}(\ket{10}\bra{10} + \ket{01}\bra{01}+\alpha\ket{10}\bra{01}+\alpha^*\ket{01}\bra{10}) . 
  \end{eqnarray}
Without loss of generality, we can take 
$\alpha$ to be real, since any phase can be absorbed into the measurement settings. In this case, non-unit visibility leads to a partially mixed heralded state of the form:
\begin{equation}
\rho_{HV} =\sqrt{V} \ketbra{\Psi}{\Psi}_H + (1 - \sqrt{V}) \rho_M,
\end{equation}
where $\rho_M = \frac{1}{2}(\ketbra{01}{01} + \ketbra{10}{10})$ represents a fully mixed state with one photon on either Alice’s or Bob’s side. This mixture allows the central station to gain partial information about the photon’s origin, thereby reducing the quality of the heralded state.

Multi-photon components are modeled via the second-order correlation function $g^{(2)}$. We represent the emitted state as a mixture of Fock states, such as
\begin{equation}
\rho_s = \sum_i P_i \rho_i, \quad \text{with} \quad \rho_s \approx P_0 \rho_0 + P_1 \rho_1 + P_2 \rho_2 + O(P_3),
\end{equation}
where $\rho_i = \ketbra{i}{i}$ denotes the $i$-photon component and $P_i$ its associated probability. For small multi-photon contributions, $g^{(2)}$ is given by
\begin{equation}
g^{(2)} = \frac{2 P_2}{(P_1 + 2 P_2)^2}.
\end{equation}
For comparing the performance of the two sources, we consider state-of-the-art values for both source types, as shown in the following table:
\begin{table}[h!]
    \centering
    \begin{tabular}{l|c|c}
        \; \textbf{Source} & \boldmath$g^{(2)}$ & \boldmath$V$ (\%) \\
        \hline \hline
        \; QD \cite{Tomm_2021,Uppu20} \; & \;  0.01 \; & \; 97.5 \;  \\
        \; SPDC  \cite{Zhao2020}     & 0.02 & 99  \\
    \end{tabular}
    \caption{Photon source parameters used in the simulations.}
    \label{tab:source_params}
\end{table}

\begin{figure}
    \centering
    \includegraphics[width=0.48\linewidth]{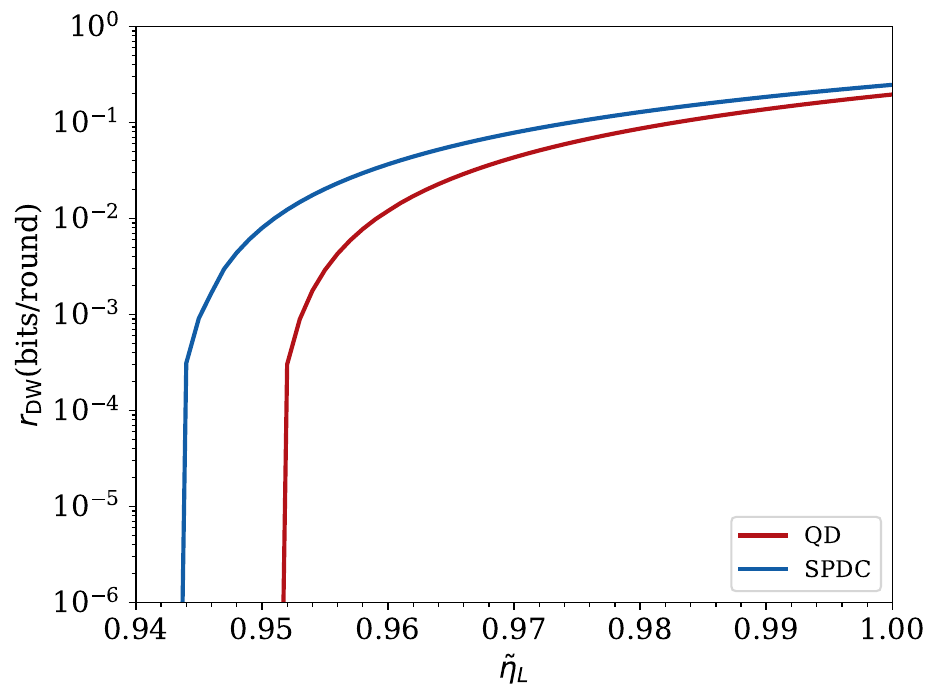}
    \includegraphics[width=0.48\textwidth]{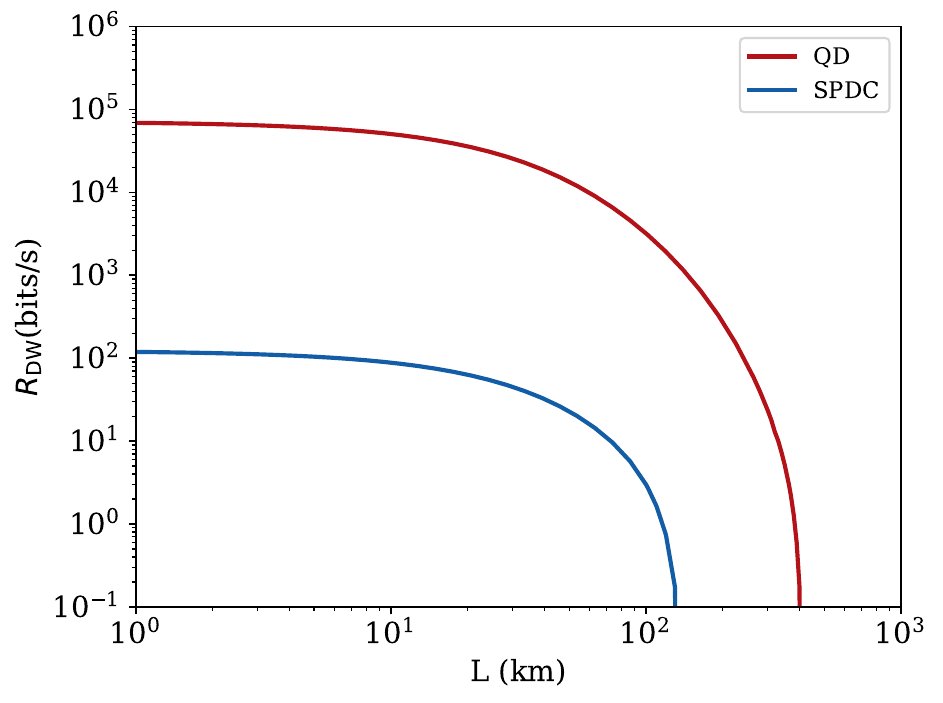}
    \caption{Comparison of DW rate and asymptotic key rate for QDs and SPDC sources. The parameters used for these plots are: $g^{(2)}=0.01$, $V=97.5\%$ for QDs; $g^{(2)}=0.02$, $V=99\%$ for SPDC; $95\%$ detector efficiency; 1 dark count/s at the central detector and $T=0.01$. In the right panel, $\tilde{\eta}_L = 98.3\%$ ($\eta_L = 93.4\%$), the single-photon generation rate for the QD is set to \SI{75}{\mega\hertz}, while the SPDC one is computed as $R_G^2$, with $R_G$ from \eqref{eq:SPDC_gen}, $\tilde{\xi}=0.1$ and $R_G=$\SI{1}{\giga\hertz}.}
    \label{fig:QD_VS_SPDC}
\end{figure}

In Figure~\ref{fig:QD_VS_SPDC} we compared the asymptotic key rate for both sources. All simulations assume detector efficiencies of $\tilde{\eta}_D=95\%$ and a dark-count rate at the central detector of 1 count/s. More details on how we model the effect of dark-counts can be found in Section \ref{sec:darkcounts}. In the left panel, we plot the Devetak-Winter rate $r_\text{DW}$ as a function of local loss $\tilde{\eta}_L$, at zero distance between Alice and Bob ($L=0$). We note that SPDC shows higher tolerance to local noise, while maintaining a positive rate down to $\tilde{\eta}_L = 94.4\%$ ($\eta_L = 89.7\%$), compared to $\tilde{\eta}_L = 95.2\%$ ($\eta_L = 90.4\%$) for QDs. In the right panel, we fix $\tilde{\eta}_L = 98.3\%$, ensuring that both sources generate an asymptotic key rate $r_\text{DW}>10^{-1}\;$bits.We then plot the asymptotic key rate $R_\text{DW}$, i.e., without accounting for finite-size effects, as a function of distance. This is defined as $R_\text{DW}=r_\text{DW}\,R_H\,\nu$, where the DW rate is multiplied by the heralding rate and the source generation rate. In this regime, QDs outperform SPDC due to their higher repetition rate. For this simulation, we set QDs to operate at \SI{75}{\mega\hertz} \cite{Tomm_2021,Uppu20,Wang19}, while SPDC sources are limited by the quality constraints imposed by $g^{(2)}$. In particular, the SPDC key rate scales with the probability of simultaneous photon generation on both sides, i.e., $\propto R_G^2$, where
    \begin{equation}
        R_G = \tilde{\xi}^2 \eta_{\text{SPDC}} R_P.\label{eq:SPDC_gen}
    \end{equation}
Here, $\tilde{\xi}$ is the squeezing parameter, $\eta_{\text{SPDC}}$ the heralding efficiency, and $R_P$ the pump rate, fixed here to \SI{1}{\giga\hertz}. Since $g^{(2)} \approx 2\xi^2$ for $g^{(2)} \ll 1$, improving source quality (reducing $g^{(2)}$) reduces generation probability, and hence the key rate.

\subsection{Impact of Parameters}
\begin{figure}
\centering
\includegraphics[width=0.5\linewidth]{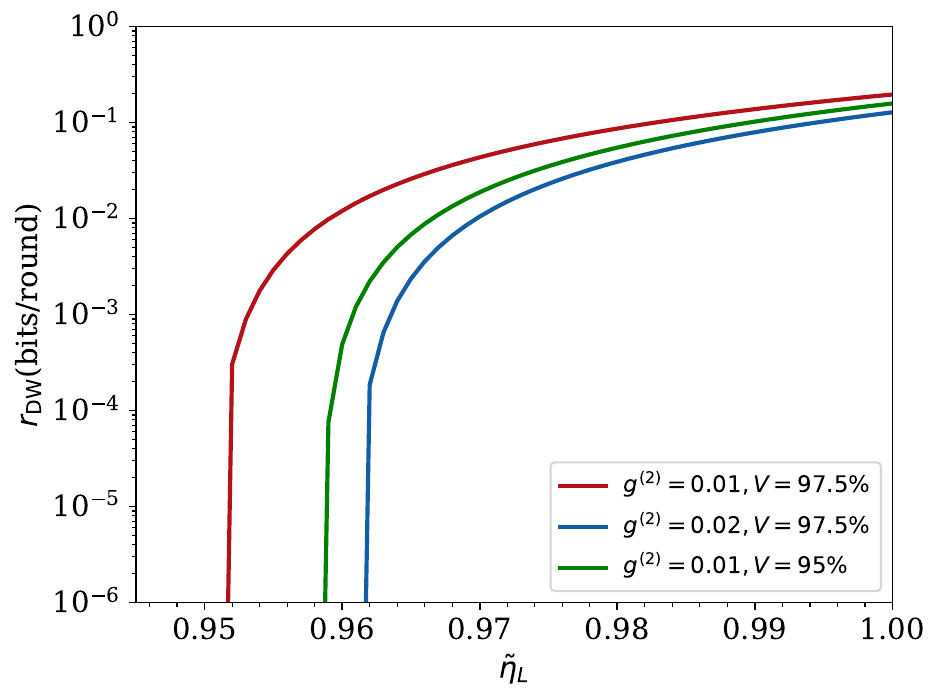}
\caption{DW key rate as a function of local efficiency $\tilde{\eta}_L$ for QDs, for different values of $V$ and $g^{(2)}$. The parameters used for these plots are: $95\%$ detector efficiency; 1 dark count/s at the central detector and $T=0.01$.}
\label{fig:QD_params}
\end{figure}

\begin{figure}[htb]
\centering
\includegraphics[width=0.5\linewidth]{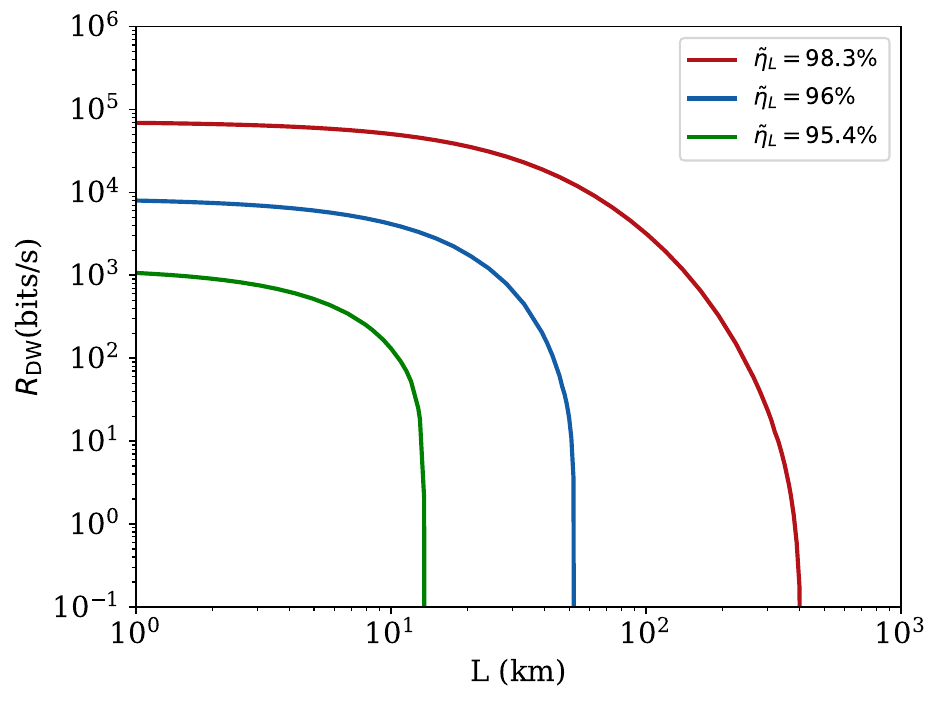}
\caption{Asymptotic key rate as a function of distance for a quantum dot source, for different values of local efficiency $\tilde{\eta}_L$. The parameters used for these plots are: $g^{(2)}=0.01$, $V=97.5\%$ for QDs; $95\%$ detector efficiency; 1 dark count/s at the central detector, $T=0.01$ and and single-photon generation rate set to \SI{75}{\mega\hertz}.}
\label{fig:QD_etas}
\end{figure}

Given the previous analysis and since this work focuses on long-distance DIQKD, we henceforth use QDs as the single-photon source for our scheme. We start by analyzing how the performance of the QD-based protocol degrades as we move away from state-of-the-art parameters. Figure~\ref{fig:QD_params} shows the DW rate as a function of $\tilde{\eta}_L$ for various values of $V$ and $g^{(2)}$. We assume detector efficiency of $95\%$ and a dark-count rate of 1 count/s (see Section \ref{sec:darkcounts}). The plots highlight how photon indistinguishability and multi-photon contribution affect the scheme's noise tolerance. We also study how variations in local efficiency affect performance as a function of distance when considering state-of-the-art parameters. In Fig.~\ref{fig:QD_etas}, we consider the state-of-the-art QD parameters and compare three values of $\tilde{\eta}_L$: $\tilde{\eta}_L = 98.3\%$ (meaning $\eta_L=93.4\%$) corresponding to an DW rate $r_\text{DW}>10^{-1}$ bits at zero distance, $\tilde{\eta}_L = 96\%$ ($\eta_L=91.2\%$), yielding $r_\text{DW} > 10^{-2}$ bits, and $\tilde{\eta}_L = 95.4\%$ ($\eta_L=90.6\%$), where $r_\text{DW} > 10^{-3}$ bits.
These values illustrate how the local imperfections impact the achievable asymptotic key rate $R_\text{DW}$ and distance.

\begin{figure}[htb]
\centering
\includegraphics[width=0.5\linewidth]{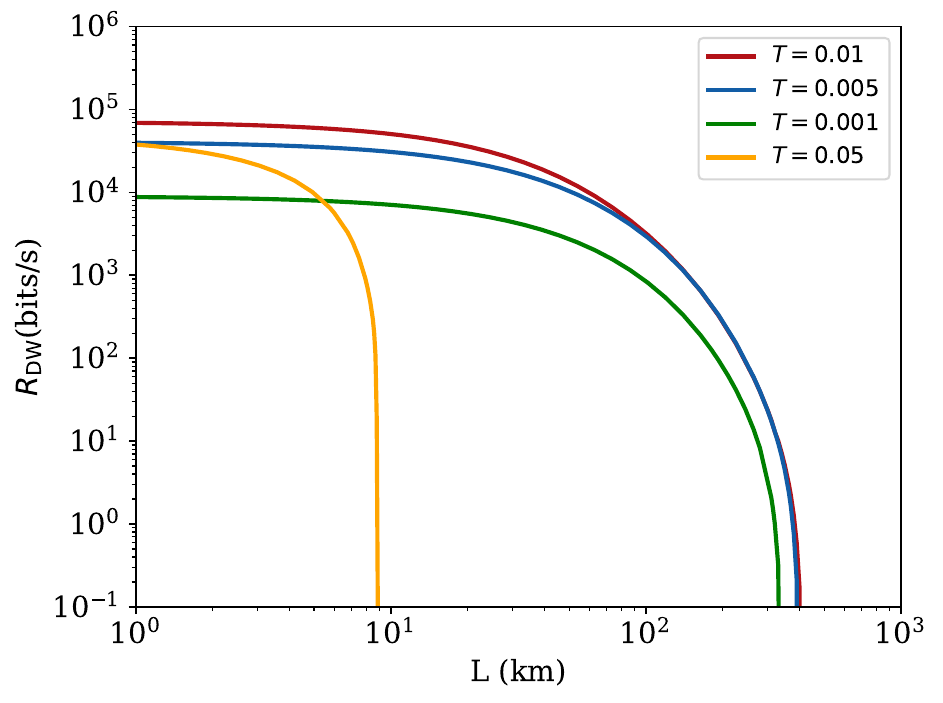}
\caption{Asymptotic key rate as a function of distance for a quantum dot source, for different values of $T$. The parameters used for these plots are: $g^{(2)}=0.01$, $V=97.5\%$ for QDs; $95\%$ detector efficiency; 1 dark count/s at the central detector and single-photon generation rate set to \SI{75}{\mega\hertz}.}
\label{fig:QD_T}
\end{figure}

Finally, to determine the optimal value of the transmittance $T$, we compare the asymptotic key rate as a function of distance $L$ for different values of $T$, as shown in Fig.~\ref{fig:QD_T}. For this simulation, we used state-of-the-art parameters for the quantum dot, $\tilde{\eta}_L = 98.3\%$ (corresponding to $\eta_L = 93.4\%$), a detection efficiency of $95\%$ for each detector, one dark count per second at the central station, and a generation rate of \SI{75}{\mega\hertz}. Larger values of $T$ increase the heralding rate and thus the key rate, but they also degrade the quality of the state. This trade-off can already reduce the key rate at short distances and, more significantly, makes the protocol more sensitive to dark counts, limiting the maximum achievable distance. Conversely, smaller values of $T$ yield higher-quality states but reduce the heralding rate and, therefore, the key rate. We find that $T = 0.01$ provides the best performance for long-distance DIQKD.

\section{Dark Counts} \label{sec:darkcounts}
In this section, we discuss the effect of dark counts on the produced state. First, note that only dark counts occurring at the heralding station are relevant, as they constitute the dominant source of false heralding events in long-distance implementations. Dark counts at other locations, such as Alice’s or Bob’s measurement stations or local source heralding detectors in the case of using SPDC, are negligible in practice. This is because only heralded events are retained in the protocol, and the probability that an accidental dark count at a local station results in a valid heralded event is proportional to $p_\text{dark}^2$, which is negligible. For instance, if a dark count at a source detector coincides with a true detection at the central station, the event occurs with probability proportional to $p_{\text{dark}} \, T \, 10^{-\alpha L/20}$. If both the source and central station simultaneously produce dark counts, the probability is of order $p_{\text{dark}}^2 \, (1 - T)$, which is even smaller. These are negligible compared to the correct heralding probability $T \, 10^{-\alpha L/20}$.  
Similarly, dark counts at the measurement stations have a negligible effect, since they only contribute to the statistics if they coincide with a successful heralding event. However, the joint probability of both events occurring is given by the product of the heralding probability and $p_\text{dark}$, which remains negligible compared to the heralding probability itself.

In contrast, dark counts at the central heralding station can directly trigger false heralding events, especially when the correct heralding rate is low due to high channel losses at long distances. We model this effect as follows: suppose a photon pair is emitted, and both photons are successfully directed toward Alice’s and Bob’s measurement stations. The probability of this is approximately $(1 - T)^2 \approx 1$, since $T \sim 10^{-2}$, and terms of order $(1 - T) T$ are neglected. A dark count then falsely heralds the state $\ket{11}$. This state undergoes local loss channels on Alice’s and Bob’s sides, characterized by a transmission efficiency $\tilde{\eta}_L$, resulting in the mixed state
\begin{equation}
\rho_{\text{HDC}} = \tilde{\eta}_L^2 \ketbra{11}{11} + \tilde{\eta}_L (1 - \tilde{\eta}_L) \left( \ketbra{01}{01} + \ketbra{10}{10} \right) + (1 - \tilde{\eta}_L)^2 \ketbra{00}{00}.
\end{equation}

The desired state $\rho_H$ is successfully heralded with rate $R_H$, while false heralding events due to dark counts at the central station occur at rate $R_{dc}$, leading to the state $\rho_{\text{HDC}}$. The overall state is then given by
\begin{equation}
\rho = \frac{1}{R_H + R_{dc}} \left( R_H \rho_H + R_{dc} \rho_{\text{HDC}} \right).
\end{equation}
This model has been implemented in the previous numerical simulations and also in \cite{git}.


\section{Finite-size analysis}\label{App:finite_size}
This section presents the finite size analysis performed in this work. We define all relevant functions and parameters necessary to reproduce our results, following closely the framework explained and rigorously proved in detail in \cite{Nadlinger_2022}. We do not provide a new security proof or additional theoretical developments, as these are comprehensively covered in \cite{Nadlinger_2022}. Instead, we restate key definitions and procedures here solely for the sake of clarity and reproducibility of our numerical findings.

We begin by formally defining the completeness and soundness parameters that characterize the security of the protocol~\cite{Tan2022}. In the a protocol, the honest parties, Alice and Bob, may either abort or accept and generate keys $K_A$ and $K_B$ of length $\ell$. The protocol is said to be $\varepsilon_{\text{com}}$-complete if, when all devices behave honestly, the probability of aborting is at most $\varepsilon_{\text{com}}$. It is $\varepsilon_{\text{{sound}}}$-sound  if, for any implementation of the protocol,
\begin{equation}
    P(accept)\frac{1}{2}\left\lVert \sigma_{K_A K_B E'}-\left(\frac{1}{2^\ell}\sum_k\ket{kk}\bra{kk}_{K_A K_B}\right)\otimes\sigma_{E'}\right\rVert_1 \leq \varepsilon_{\text{sound}},
\end{equation}
where $\sigma$ is the state conditioned on the protocol being accepted, and $E'$ denotes all side information available to an eavesdropper, Eve, at the end of the protocol. 

In the following, we define the functions that have been used for computing the key rate for finite-size statistics. We start by writing the CHSH-based entropy bound \cite{Noisy_Preprocessing_Ho20}
 
\begin{equation}
    \eta(s) = \begin{cases}
        0 & \text{if}\; \omega\in\mathcal{C}\\
        1 - h\left(\frac{1+\sqrt{16 \omega^2 - 16 \omega+3}}{2}\right) + h\left(\frac{1+\sqrt{1-q_n(1-q_n)(8-(8\omega-4)^2))}}{2}\right) & \text{elif}\; \omega\in\mathcal{Q}\\
        \text{undefined}  & \text{else},
    \end{cases}
\end{equation}

where $q_n$ is the bit-flip probability of the noisy preprocessing, $\mathcal{C}=[\frac{1}{4},\frac{3}{4}]$, $\mathcal{Q}=[\omega_\text{min},\omega_\text{max}]$, with $\omega_\text{min}=\frac{1-1/\sqrt{2}}{2}$ and $\omega_\text{max}=\frac{1+1/\sqrt{2}}{2}$, and $h(X)$ represents the binary entropy. Then we can express the family of linear lower bounds on the entropy
 
\begin{equation}
    g_t(\omega) = \eta(t) + (\omega - t) \frac{d\eta(t)}{d t},
\end{equation}
 
from which we obtain
 
\begin{equation}
    f_t(\delta_u)= \begin{cases}
        \frac{1}{\gamma} g_t(0) + \left(1-\frac{1}{\gamma}\right) g_t(1) &\text{if} \;u=0\\
        g_t(1) &\text{if}\; u=1 \;\text{or}\; u=\perp,
    \end{cases}
\end{equation}
 
where $\delta_u$ are three delta distributions over $\{0,1,\perp\}$, with $\perp$ indicating a key-round. If we introduce a probability distribution $p:{0,1,\perp}\rightarrow\mathds{R}$, we can define the following family of functions
 
\begin{equation}
    f_t(p) = \sum_{u=0,1,\perp} p(u) f_t(\delta_u),
\end{equation}
 
with variance equal to
 
\begin{equation}
    \text{Var}_p(f_t) = \sum_{u=0,1,\perp} p(u) f_t(\delta_u)^2 - f_t(p)^2.
\end{equation}
 
To compute the key rate, we also need the following functions
 
\begin{align}
    \theta_\varepsilon &= \log\frac{1}{1-\sqrt{1-\varepsilon^2}}\leq \log\frac{2}{\varepsilon^2},\\
    K_{\alpha'}(f_t) &=\frac{1}{6(2-\alpha')^3\ln 2}2^{(\alpha'-1)(2+g_t(1)-g_t(\omega_\text{min}))}\ln^3(2^{2+g_t(1)-g_t(\omega_\text{min})}+e^2).
\end{align}
 
We will employ the upper bound for $\theta_\varepsilon$, as its precise value can render numerical computations unstable. By considering the probability distribution $q(\omega) = (\gamma(1-\omega),\gamma\omega,1-\gamma)$, we can write
 
\begin{align}
    \Delta(f_t,\omega) &= \eta(\omega) - f_t(q(\omega)) = \left(\eta(\omega)-\eta(t)\right) - \frac{d \eta(t)}{d t} \left(\omega-t\right),\\
    V(f_t,q(\omega)) &= \frac{\ln2}{2}\left(\log33+\sqrt{2+\text{Var}_{q(\omega)}(f_t)}\right)^2.    
\end{align}
 
Observe that
 
\begin{align}
    \inf_{\omega\in\mathcal{Q}}\Delta(f_t,\omega) &= \Delta(f_t,t) \equiv 0,\\
    \max_{\omega\in\mathcal{Q}} \text{Var}_{q(\omega)}(f_t)&\leq\frac{2+\sqrt{2}}{4\gamma}\left(g_t(1) - g_t(0)\right)^2.
\end{align}
 
We will also need to define
 
\begin{equation}
    Y_b(x) = b W\left(\frac{e^{x/b}}{b}\right),
\end{equation}
 
where $W$ is the principal branch of the Lambert function and $b=\frac{4}{\ln 2}$. Finally, we introduce the parameters:
 
\begin{align}
\begin{split}
    n\in\mathds{N} &\;\;\text{number of rounds}\\
    \gamma\in(0,1) &\;\;\text{probability of a test-round}\\
    \omega_{thr} &\;\;\text{threshold CHSH winning probability}\\
    m &\;\;\text{length of the error correction syndrome}\\
    t \in \left(\frac{3}{4},\frac{1+1/\sqrt{2}}{2}\right] &\;\;\text{CHSH winning probability}\\
    \varepsilon_h>0 &\;\;\text{hashing collision probability}\\
    \varepsilon_{PA}>0 &\;\;\text{privacy amplification parameter}\\
    \varepsilon_{EA}>0 &\;\;\text{entropy accumulation parameter}\\
    \alpha' \in (1,2)&\;\;\text{Renyi parameter}\\
    \alpha'' \in \left(1,1+\frac{1}{\log5}\right)&\;\;\text{Renyi parameter}.
\end{split}
\end{align}
 
Specifying the smoothing parameters $\varepsilon_s,\varepsilon_s',\varepsilon_s''>0$ such that
 
\begin{equation}
    \varepsilon_s'+2\varepsilon_s''<\varepsilon_s.
\end{equation}
 
Utilizing the parameters and functions delineated so far, it can be shown (see \cite{Nadlinger_2022}) that our protocol generates a ($\max(\varepsilon_{EA},\varepsilon_{PA}+2\varepsilon_{s}$)+4$\varepsilon_h$)-sound key, with a length of

\begin{equation}\label{key}
    \ell = Y_b(l),
\end{equation}
 
with 
 
\begin{align}\label{arg}
\begin{split}
    l\, = \, & n \; g_t(\omega_{thr}) + n \inf_{\omega\in\mathcal{Q}} \Delta(f_t,\omega) - n(\alpha' -1)\,  \max_{\omega\in\mathcal{Q}}V(f_t,q(\omega)) \\
    & - n\,(\alpha'-1)^2 \, K_{\alpha'}(f_t) - n \,\gamma - n\,(\alpha''-1) \,\log^2(5)\\
    & - \frac{1}{\alpha'-1}\left(\theta_{\varepsilon_s'}+\alpha'\,\log(\frac{1}{\varepsilon_{EA}})\right)- \frac{1}{\alpha''-1} \left(\theta_{\varepsilon_s''}+\alpha''\,\log(\frac{1}{\varepsilon_{EA}})\right)\\
    & - 3\,\theta_{\varepsilon_s-\varepsilon_s'-2\varepsilon_s''}- 5\log(\varepsilon_{PA})-m-264.
\end{split}
\end{align}

Given that the values of $\varepsilon_s, \varepsilon_s', \varepsilon_s'', \varepsilon_{EA}, \varepsilon_{PA}$ have been set to $(10^{-10}, 3\cdot 10^{-11}, 3\cdot 10^{-11}, 10^{-10}, 10^{-10})$, and $\varepsilon_h=2^{-61}$ is determined through the use of the VHASH algorithm, the protocol is proved to be $3\cdot 10^{-10}$ sound. All the rest of the parameters, $\alpha', \alpha'', t,\gamma, q_n$ have been optimized to maximize $\ell$.
The value of the threshold CHSH winning probability $\omega_{thr}$ and the length of the error correction syndrome $m$ were chosen according to those suggested in \cite{Nadlinger_2022}.
Specifically, we employed
 
\begin{align}
\omega_{thr}&=1-\frac{q_{thr}}{\gamma}\\
q_{thr}&= q + k\sqrt{\frac{q(1-q)}{n}}
\end{align}
 
with $k=3$, $q=\gamma(1-\omega)$, $\omega=\frac{4+S}{8}$, and $S$ representing the CHSH-score.
This selection ensures that the completeness of the protocol is $\varepsilon_{com}<0.01$.
As for the error correction syndrome we used
 
\begin{equation}
    m = n ((1-\gamma) H(A_1\vert B_3)+\gamma h((4-S)/8))+ 50\sqrt{n}.
\end{equation}
 
The CHSH-score $S$ and $H(A\vert B)$ have been computed using the optimal measurements described in the Supplementary Material \ref{App:measurements} for a fixed overall local efficiency $\eta_L$. 
For computing the key rate we have imposed our protocol to be $10^{-2}$-complete and $3\cdot 10^{-10}$-sound.

\bibliography{note}